\documentclass[superscriptaddress,twocolumn,pre,nofootinbib]{revtex4}

\usepackage{amsmath}
\usepackage{graphicx,color}

\newcommand{\be}{\begin{equation}}
\newcommand{\ee}{\end{equation}}
\newcommand{\bea}{\begin{eqnarray}}
\newcommand{\eea}{\end{eqnarray}}

\begin{document}
\sloppy


\title{Random transitions described by the stochastic Smoluchowski-Poisson system \\
and by the stochastic Keller-Segel model}

\author{P.H. Chavanis and L. Delfini}
\affiliation{Laboratoire de Physique Th\'eorique (UMR 5152), Universit\'e Paul Sabatier, IRSAMC, 118 route de Narbonne, 31062 Toulouse cedex 4.}

\begin{abstract}
We study random transitions between two metastable states that appear below a
critical temperature in a one dimensional self-gravitating Brownian gas
with a
modified Poisson equation experiencing a second order phase transition from a
homogeneous phase to an inhomogeneous phase [P.H. Chavanis and L. Delfini, Phys.
Rev. E {\bf 81}, 051103 (2010)]. We numerically solve the $N$-body Langevin
equations and the stochastic Smoluchowski-Poisson system which takes
fluctuations (finite $N$ effects) into account. The system switches back and
forth between the two metastable states (bistability) and the particles
accumulate successively at the center or at the boundary of the domain. We
explicitly show that these random transitions exhibit the phenomenology of the
ordinary Kramers problem for a Brownian particle in a double-well potential. The
distribution of the residence time is Poissonian and the average lifetime of a
metastable state is given by the Arrhenius law, i.e. it is proportional to the
exponential of the barrier of free energy $\Delta F$  divided by the energy of
thermal excitation $k_B T$. Since the free energy is proportional to the number
of particles $N$ for a system with long-range interactions, the lifetime of
metastable states scales as $e^N$ and is considerable for $N\gg 1$. As a result,
in many applications, metastable states of systems with long-range interactions
can be considered as stable states. However, for moderate values of $N$, or
close
to a critical point,
the lifetime of the metastable states is reduced since the barrier of free
energy decreases. In that case, the fluctuations become important and the mean
field approximation is no more valid. This is the situation considered in this
paper. By an
appropriate change of notations, our results also apply to bacterial populations
experiencing chemotaxis in biology. Their dynamics can be described by a
stochastic Keller-Segel model that takes fluctuations into account and goes
beyond
the usual mean field approximation.

\end{abstract}

\maketitle


\section{Introduction}

The theory of Brownian motion is an important topic in physics \cite{risken}.
In most studies, the Brownian particles do not interact, or have short-range
interactions. When an overdamped Brownian particle evolves in a double-well
potential $V(x)$ (i.e. a potential with two minima and a maximum), it undergoes
random transitions between the minima of the potential (metastable states). The
position $x(t)$ of the particle switches back and forth between the location of
the minima and presents a  phenomenon of  bistability.  The distribution
$P(x,t)$ of the position of the particle is governed by a Fokker-Planck equation
called the  Smoluchowski equation. At  equilibrium, we get the Boltzmann
distribution $P(x)=Z(\beta)^{-1}e^{-\beta V(x)}$ which is bimodal. The
transition probability between the two minima, or the escape time, has been
determined by Kramers in a famous paper \cite{kramers,hanggi}. It is given by
the Arrhenius law $e^{\Delta V/k_B T}$ where $\Delta V=V_{max}-V_{min}$ is the
difference of potential between the metastable state (minimum) and the unstable
state (maximum). Furthermore, the distribution of the residence time of the
system in the metastable states is Poissonian.

The study of Brownian particles with long-range interactions (corresponding to the canonical ensemble) is a challenging problem \cite{hb1,hb2,hb5,longshort,initialvalue}. A system of fundamental interest is the self-gravitating Brownian gas model studied in \cite{sc}. This model may be relevant to describe the dynamics of dust particles in the solar nebula (where the particles experience a friction with the gas and a stochastic force due to small-scale turbulence) and the formation of planetesimals by gravitational collapse \cite{aaplanete}. In the strong friction limit $\xi\rightarrow +\infty$, and in the thermodynamic limit $N\rightarrow +\infty$, the evolution of the density of the self-gravitating Brownian gas is described by the Smoluchowski equation coupled to the Poisson equation. In a space of dimension $d\ge 2$, these equations display a phenomenon of {\it isothermal collapse} \cite{sc} below a critical temperature $T_c$  leading to the formation of a Dirac peak \cite{post}. By contrast, in $d=1$, there is no collapse and the density always reaches a stable steady state \cite{sc}.

Interestingly, the Smoluchowski-Poisson (SP) system is closely related to the
Keller-Segel (KS) model that describes the chemotaxis of bacterial populations
in biology \cite{ks}. In this model, the bacteria undergo Brownian motion
(diffusion) but they also secrete a chemical substance (a sort of pheromone) and
are collectively attracted by it. It turns out that this long-range interaction
is similar to the gravitational interaction in astrophysics \cite{jeanstype,tcrit}. As
a result, the KS model displays a phenomenon of {\it chemotactic collapse} in
$d\ge 2$ leading to Dirac peaks. Actually, the KS model is more general than the
SP system because the Poisson equation is replaced by a reaction-diffusion
equation. In certain approximations (no degradation and large diffusivity of the
secreted chemical) the reaction-diffusion equation can be reduced to a modified
Poisson equation that includes
a sort of ``neutralizing background'' (similar to the Jellium
model of a plasma). As a result, a spatially homogeneous distribution of
particles is always a steady state of the KS model, which is not the case for
the ordinary SP system.\footnote{In astrophysics, when we consider the dynamical
stability of an infinite homogeneous self-gravitating system, we have
to do the so-called ``Jeans swindle''
\cite{bt}. There is no such swindle in the context of chemotaxis
\cite{jeanstype}.} This analogy prompts us to consider a
model of self-gravitating Brownian particles with a modified Poisson equation
\cite{cd}. By an appropriate change of notations, this ``gravitational'' model
can be mapped onto a ``chemotactic'' model. In that model, a spatially
homogeneous phase exists at any temperature but it becomes unstable below a
critical temperature $T_c^*$ where it is replaced by an inhomogeneous
(clustered) phase. This second order phase transition exists in any dimension of
space including the dimension $d=1$. This is interesting because the usual SP
system does not present any phase transition in $d=1$ \cite{sc}. In $d\ge 2$,
this second order phase transition at $T_c^*$ adds to the isothermal collapse at
$T_c$ described above. A
detailed study of phase transitions in this model has been performed in
\cite{cd}  in various dimensions of space.

The inhomogeneous (clustered) phase that appears below $T_c^*$ is degenerate.
The particles may accumulate either at the center of the domain ($r=0$) or at
the boundary of the domain ($r=R$). These two configurations correspond to local
minima of the mean field free energy $F[\rho]$ at fixed mass,
where $\rho({\bf r})$ is the  density field. Since the phase transition is
second order, these two minima are at the same hight. On the other hand, the
unstable homogeneous phase $\rho=M/V$ corresponds to a saddle point of free
energy. In the $N\rightarrow +\infty$ limit, the evolution of the system is
described by the deterministic SP system (with the modified Poisson equation).
For $T<T_c^*$ this  equation relaxes towards one of the two metastable states
(the choice depends on the initial condition and on a notion of basin of
attraction) and remains in that state for ever. However, if we take finite $N$
effects into account, there are fluctuations, and the system undergoes
random transitions between the two metastable states. These fluctuations are
particularly important close to the critical point $T_c^*$ where the barrier of
free energy $\Delta F$ to overcome is small.\footnote{We expect 
similar results in the microcanonical ensemble for isolated self-gravitating
systems with a modified Poisson equation. In that case, two degenerate
metastable states appear below a critical energy $E_c^*$ \cite{cd}. The system
undergoes random transitions between these metastable states that are controlled
by the barrier of entropy $\Delta S$ instead of the barrier of free
energy $\Delta F$. For Hamiltonian self-gravitating systems the noise is due
only to finite $N$ effects while for self-gravitating Brownian particles it is
due both to the stochastic force and to finite $N$ effects.  Furthermore, before
reaching the statistical equilibrium state (and undergoing random transitions),
isolated self-gravitating systems may be stuck in non-Bolzmannian quasi
stationary states (QSSs) \cite{teles1,teles2,joyce} while self-gravitating
Brownian particles in the overdamped limit directly relax towards the Boltzmann
statistical equilibrium state (and undergo random transitions) without forming
QSSs. Finally, the dynamical behavior of isolated self-gravitating systems with
the usual Poisson equation is very different from their evolution with the
modified Poisson equation since there is no homogeneous phase in that case
\cite{ijmpb}. In
$d\ge 3$, the system develops a gravothermal catastrophe below a critical energy
$E_c$ leading to a binary star surrounded by a hot halo. In $d=1$ and $d=2$
there is no collapse and the system reaches an inhomogeneous statistical
equilibrium state.}

In this paper, we study these random transitions by numerically solving the
$N$-body Langevin equations and the stochastic SP  system in $d=1$ (with the
modified Poisson equation). The stochastic SP system takes fluctuations into
account by including a noise term whose strength is proportional to
$1/\sqrt{N}$. We investigate the random transitions between the two
metastable states and show that they follow the phenomenology of the Kramers
problem for a Brownian particle in a double-well potential. The equilibrium
distribution of the density is $P[\rho]=Z(\beta)^{-1}e^{-\beta F[\rho]}$ which
is bimodal. The system switches
back and forth between the two metastable states (bistability) and the particles
accumulate successively at the center or at the boundary of the domain. The
difference with the Kramers problem is that our stochastic variable is a density
field $\rho(x,t)$ instead of the position $x(t)$ of a particle. Furthermore, in
our problem, the $N$ particles interact collectively and create their own free
energy
landscape while in the usual Kramers
problem a unique particle (or a set of non-interacting particles) moves in an
externally imposed potential $V(x)$. These analogies and differences with the
Kramers problem make the present model interesting to study. We show that the
distribution of the residence time $P(\tau)$ is Poissonian and that the average
time $\langle\tau\rangle$ spent by the system in a metastable state is given by
the Arrhenius law $e^{\Delta F/k_B T}$ where $\Delta F=F_{saddle}-F_{meta}$ is
the
difference of free energy between the metastable state (minimum) and the
unstable state (saddle point). Since the free energy $F$ scales as $N$ for systems
with long-range interactions, this implies that the lifetime of the metastable
states generically scales as $e^N$ except close to a critical point where the
barrier of free energy $\Delta F$ is small (this result was previously
reported in \cite{rieutord,art,metastable}). Therefore, when $N\gg 1$ the
metastable states have considerably long lifetimes and they may be considered as
{\it stable} states in practice.\footnote{This result has important consequences
in astrophysics where the number of stars in a globular cluster may be as large
as $10^6$ \cite{bt}, resulting in a lifetime of the metastable states of the
order of $e^{10^6}t_D$ where $t_D$ is the dynamical time. The probability to
cross the barrier of entropy is exponentially small since it requires very
particular correlations. As a result, self-gravitating systems may be found in
long-lived metastable states (local entropy maxima) although there is no
statistical equilibrium state (global entropy maximum) in a strict sense
(see \cite{metastable,ijmpb} for more detail). In reality, the lifetime of globular clusters is ultimately controlled by the processes of evaporation and core collapse \cite{bt}.}
The probability to pass from one metastable state to the other is a rare even since it scales as $e^{-N}$. Random transitions between metastable states
can be seen only close to the critical point, for moderate
values of $N$, and for sufficiently long times. This
result also implies that the limits $N\rightarrow +\infty$ and $T\rightarrow
T_c^*$ do not commute. Indeed, close to the critical point, the fluctuations
become important and the mean field approximation is no more valid (or requires
a larger and larger number of particles as $T\rightarrow T_c^*$).

The paper is organized as follows. In Sec. \ref{sec_obmf} we review the
basic equations describing a gas of  Brownian particles with long-range
interactions in the overdamped limit. The equation of main interest is the
stochastic Smoluchowski equation (\ref{sto5}) that takes fluctuations (finite
$N$ effects) into account. In the $N\rightarrow +\infty$ limit, the fluctuations
become negligible and we recover the mean field Smoluchowski equation
(\ref{h13}). In Sec. \ref{sec_formal}, we apply these equations to
self-gravitating Brownian particles and bacterial populations with a modified
Poisson equation. In Sec. \ref{sec_ssp} we specifically consider the dimension
$d=1$ where a second order phase transition between a homogeneous phase and an
inhomogeneous (clustered phase) appears below a critical temperature $T_c^*$. We
numerically solve the modified stochastic Smoluchowski-Poisson system
(\ref{efd1})-(\ref{efd2}) and describe random changes between the two metastable
states below $T_c^*$. We characterize these transitions and show that they
display a phenomenology similar to the standard Kramers problem. We also present
results from direct $N$-body simulations.

\section{Overdamped Brownian particles with long-range interactions}
\label{sec_obmf}

\subsection{The Langevin equations}
\label{sec_lang}

We consider a system of $N$ Brownian particles in interaction. The
dynamics of these particles is governed by the coupled stochastic
Langevin equations
\begin{eqnarray}
{d{\bf r}_{i}\over dt}&=&{\bf v}_{i},
\qquad\qquad\qquad\qquad\qquad\qquad\qquad\quad\nonumber\\
{d{\bf v}_{i}\over dt}&=&-\frac{1}{m}\nabla_i U({\bf r}_{1},...,{\bf r}_{N})-\xi
{\bf v}_{i}+\sqrt{2D}{\bf R}_{i}(t).
\label{bkram1}
\end{eqnarray}
The particles interact through
the potential $U({\bf r}_{1},...,{\bf
r}_{N})=\sum_{i<j}m^2 u(|{\bf r}_{i}-{\bf r}_{j}|)$. The
Hamiltonian is $H=\sum_{i=1}^{N}m{v_{i}^{2}/2}+U({\bf
r}_{1},...,{\bf r}_{N})$.  ${\bf
R}_i(t)$ is a Gaussian white noise satisfying $\langle {\bf
R}_i(t)\rangle={\bf 0}$ and $\langle
R_i^{\alpha}(t)R_j^{\beta}(t')\rangle=\delta_{ij}\delta_{\alpha\beta}
\delta(t-t')$ where $i=1,...,N$ label the particles and $\alpha=1,...,d$ the
coordinates of space. $D$
and $\xi$ are respectively the diffusion and friction
coefficients. The former measures the strength of the noise, whereas
the latter quantifies the dissipation to the external environment. We
assume that these two effects have the same physical origin, like when
the system interacts with a heat bath. In particular, we suppose that
the diffusion and friction coefficients  satisfy the Einstein relation
\begin{equation}
D=\frac{\xi k_B T}{m},
\label{einstein1}
\end{equation}
where $T$ is the temperature of the bath. The temperature measures the strength
of the stochastic force (for a given friction coefficient). For $\xi=D=0$, we
recover the
Hamiltonian equations of particles in interaction which conserve the
energy $E=H$.

\subsection{The strong friction limit}
\label{sec_overlang}

In the strong friction limit $\xi\rightarrow +\infty$, the inertia of
the particles can be neglected. This corresponds to the overdamped Brownian
model. In this paper, we restrict ourselves to this model.\footnote{The inertial Brownian model, and its connection to the overdamped Brownian model, is further discussed in \cite{initialvalue}.} The stochastic Langevin equations (\ref{bkram1}) reduce to
\begin{eqnarray}
{d{\bf r}_{i}\over dt}=-\mu \nabla_iU({\bf r}_{1},...,{\bf
r}_{N})+\sqrt{2D_{*}}{\bf R}_{i}(t),
\label{smf1}
\end{eqnarray}
where $\mu=1/(\xi m)$ is the mobility and $D_{*}=D/\xi^{2}$ is the
diffusion coefficient in physical space. The Einstein relation
(\ref{einstein1}) may be rewritten as
\begin{eqnarray}
D_{*}=\frac{k_B T}{\xi m}=\mu k_B T.
\label{einstein2}
\end{eqnarray}
The temperature measures the strength of the stochastic force (for a given
mobility).

\subsection{The $N$-body Smoluchowski equation}
\label{sec_smf}

The evolution of the
$N$-body distribution function $P_N({\bf r}_1,...,{\bf r}_N,t)$ is governed by
the $N$-body
Fokker-Planck equation \cite{hb2}:
\begin{eqnarray}
{\partial P_{N}\over\partial t}=\sum_{i=1}^{N}{\partial\over\partial {\bf
r}_{i}}\cdot \biggl\lbrack D_{*}{\partial P_{N}\over\partial{\bf r}_{i}}+\mu
P_{N}{\partial\over\partial {\bf r}_{i}}U({\bf r}_{1},...,{\bf
r}_{N})\biggr\rbrack.
\label{smf3}
\end{eqnarray}
This is the so-called $N$-body Smoluchowski equation. It can be derived directly
from the stochastic equations (\ref{smf1}). The $N$-body Smoluchowski equation satisfies an H-theorem for the free energy
\begin{eqnarray}
\label{smf5}
F[P_N]=\int P_N U\, d{\bf r}_1...d{\bf r}_N
+k_BT\int P_N\ln P_N\, d{\bf
r}_1...d{\bf r}_N\nonumber\\
-\frac{d}{2}Nk_B T\ln\left (\frac{2\pi k_B T}{m}\right ).\qquad
\end{eqnarray}
A simple calculation gives
\begin{equation}
\label{smf6}
\dot F=-\sum_{i=1}^{N}\int \frac{m}{\xi P_N}\left (\frac{k_B T}{m}{\partial
P_{N}\over\partial{\bf r}_{i}}+\frac{1}{m} P_{N}{\partial U\over\partial {\bf
r}_{i}}\right )^2\, d{\bf r}_1...d{\bf r}_N.
\end{equation}
Therefore, $\dot F\le 0$ and $\dot F=0$ if, and only if, $P_N$ is the canonical
distribution in physical space defined by Eq. (\ref{cano1}) below. Because of the
$H$-theorem, the system converges towards the canonical distribution
(\ref{cano1}) for $t\rightarrow +\infty$.

We note that the free energy may be written as $F[P_N]=E[P_N]-T S[P_N]$ where
\begin{eqnarray}
\label{smf5b}
E[P_N]=\frac{d}{2}Nk_B T+\int P_N U\, d{\bf r}_1...d{\bf r}_N,
\end{eqnarray}
\begin{eqnarray}
\label{smf5c}
S[P_N]=-k_B\int P_N\ln P_N\, d{\bf
r}_1...d{\bf r}_N\nonumber\\
+\frac{d}{2}Nk_B \ln\left (\frac{2\pi k_B T}{m}\right )+\frac{d}{2}Nk_B
\end{eqnarray}
are the energy and the entropy \cite{initialvalue}.

\subsection{The canonical distribution}
\label{sec_cano}

The statistical equilibrium state of the Brownian particles in interaction is described by the
canonical distribution \cite{initialvalue}:
\begin{eqnarray}
P_{N}({\bf r}_{1},...,{\bf r}_{N})={1\over Z(\beta)}\left (\frac{2\pi}{\beta m}\right )^{dN/2}e^{-\beta U({\bf
r}_{1},...,{\bf r}_{N})},
\label{cano1}
\end{eqnarray}
where
\begin{eqnarray}
Z(\beta)=\left (\frac{2\pi}{\beta m}\right )^{dN/2}\int e^{-\beta U({\bf
r}_1,...,{\bf r}_N)}\, d{\bf r}_1...d{\bf r}_N
\label{cano2}
\end{eqnarray}
is the partition function determined by the normalization
condition $\int P_N \, d{\bf r}_1....d{\bf r}_N=1$.
The canonical distribution (\ref{cano1}) is the steady state of the $N$-body Smoluchowski equation
(\ref{smf3}) provided that the Einstein relation (\ref{einstein2}) is satisfied. It gives the probability density of the microstate  $\lbrace{\bf r}_1,...,{\bf r}_N\rbrace$.
The free energy is defined  by $F(T)=-k_B T\ln Z(T)$. We note that the canonical
distribution (\ref{cano1}) is the minimum of $F[P_N]$
respecting the normalization condition. At equilibrium, we have  $F[P_N]=-k_B
T\ln Z(T)=F(T)$.

\subsection{The Yvon-Born-Green (YBG) hierarchy}
\label{sec_ybg}

We introduce the reduced probability distributions
\begin{equation}
\label{ybg1}
P_{j}({\bf r}_{1},...,{\bf r}_{j})=\int P_{N}({\bf r}_{1},...,{\bf r}_{N})\, d{\bf r}_{j+1}...d{\bf r}_{N}.
\end{equation}
Differentiating the defining relation (\ref{ybg1}) for $P_j$ and using Eq. (\ref{cano1}), we obtain
 the YBG  hierarchy of equations \cite{hansen,hb1,hb5}:
\begin{eqnarray}
\label{ybg2}
{\partial P_{j}\over\partial {\bf r}_{1}}({\bf r}_{1},...,{\bf r}_{j})=-\beta m^{2} P_{j}({\bf r}_{1},...,{\bf r}_{j})\sum_{i=2}^{j} {\partial u_{1,i}\over\partial {\bf r}_{1}}\nonumber\\
-\beta m^{2} (N-j)\int P_{j+1}({\bf r}_{1},...,{\bf r}_{j+1})
{\partial u_{1,j+1}\over\partial {\bf r}_{1}}d{\bf r}_{j+1},
\end{eqnarray}
where  we
have noted $u_{i,j}$ for $u(|{\bf r}_i-{\bf r}_j|)$. The first equation of the YBG hierarchy is
\begin{eqnarray}
\label{ybg3} {\partial P_{1}\over\partial {\bf r}_{1}}({\bf r}_{1})=-\beta m^{2} (N-1) \int P_{2}({\bf r}_{1},{\bf r}_{2}){\partial u_{1,2}\over\partial {\bf r}_{1}}d{\bf r}_{2},
\end{eqnarray}
where $P_{1}({\bf r}_{1})$ is the one-body distribution function and $P_{2}({\bf r}_{1},{\bf r}_{2})$ is the two-body distribution function. This equation is exact but it is not closed.

We now consider a system with long-range interactions. The proper thermodynamic limit $N\rightarrow +\infty$ amounts to
writing the Hamiltonian in the rescaled form
\begin{eqnarray}
\label{ybg4}
H=\sum_i \frac{1}{2} m v_i^2+\frac{1}{N}\sum_{i<j}m^2\tilde{u}_{ij}
\end{eqnarray}
with $r\sim t\sim v\sim m\sim \tilde{u}\sim 1$. The factor $1/N$ in front of the potential energy corresponds to the Kac scaling \cite{kac}. With this scaling $E\sim N$, $T\sim 1$, $S\sim N$, and $F\sim N$. The energy is extensive but it remains fundamentally non-additive.  For $N\rightarrow +\infty$  we can neglect the correlations between the particles \cite{cdr}. Therefore, the mean field
approximation is exact and the $N$-body distribution function can be
factorized in a product of $N$ one-body distribution functions
\begin{eqnarray}
P_N({\bf r}_1,...,{\bf r}_N)=P_1({\bf r}_1)...P_1({\bf r}_N).\label{ybg5}
\end{eqnarray}
In particular, $P_2({\bf r}_1,{\bf r}_2)=P_1({\bf r}_1)P_1({\bf r}_2)$. In that case, the first equation of the YBG hierarchy reduces to
\begin{eqnarray}
\label{ybg6} {\partial P_{1}\over\partial {\bf r}_{1}}({\bf r}_{1})=-\beta m^{2} N P_1({\bf r}_1)\int P_1({\bf r}_2) {\partial u_{1,2}\over\partial {\bf r}_{1}}d{\bf r}_{2}.
\end{eqnarray}
Introducing the average density $\rho({\bf r})=\langle
\sum_i m \delta({\bf r}-{\bf r}_i)\rangle=NmP_1({\bf r})$, we get
\begin{eqnarray}
\label{ybg7}\frac{k_B T}{m} \nabla\rho({\bf r})=-\rho({\bf r})\int \rho({\bf r}')\nabla u(|{\bf r}-{\bf r}'|)\, d{\bf r}',
\end{eqnarray}
where we used the fact that the particles are identical. Integrating this equation, we obtain the
 integral equation
\begin{eqnarray}
\rho({\bf r})=A\, e^{-\beta m \int u(|{\bf r}-{\bf r}'|)\rho({\bf r}')\, d{\bf r}'}
\label{ybg10}
\end{eqnarray}
determining the equilibrium density profile. It can be rewritten as a mean field Boltzmann distribution
\begin{eqnarray}
\label{ybg8}
\rho({\bf r})=A\, e^{-\beta m \Phi({\bf r})},
\end{eqnarray}
where $\Phi({\bf r})$ is the self-consistent mean field potential given by
\begin{eqnarray}
\Phi({\bf r})=\int u(|{\bf r}-{\bf r}'|)\rho({\bf r}')\, d{\bf r}'.
\label{ybg9}
\end{eqnarray}
The constant of integration $A$ is determined by the normalization condition $\int P_1({\bf r}_1)\, d{\bf r}_1=1$ or equivalently by the mass constraint
\begin{eqnarray}
M[\rho]=\int\rho\, d{\bf r}.
\label{ybg11}
\end{eqnarray}
We also note that, in the mean field approximation, the average potential energy is given by
\begin{eqnarray}
W[\rho]=\frac{1}{2}\int\rho\Phi\, d{\bf r}.
\label{ybg12}
\end{eqnarray}

\subsection{The distribution of the smooth density}
\label{sec_heur}

We wish to determine the equilibrium distribution of the smooth
(coarse-grained) density $\rho({\bf r})$ in position space.\footnote{The
coarse-grained density should be noted $\overline{\rho}({\bf r})$
\cite{hb5,longshort} but in order to simplify the notations we shall omit the
bar.} A {\it
microstate} is defined by the specification of the exact positions $\lbrace {\bf
r}_i\rbrace$ of the $N$ particles. A
{\it macrostate} is defined by  the specification of the density
$\rho({\bf r})$ of particles in each cell $[{\bf r},{\bf r}+d{\bf r}]$ irrespectively of their  precise position in the cell.  Let us call $\Omega[\rho]$ the unconditional number of  microstates $\lbrace{\bf r}_i\rbrace$ corresponding to the macrostate $\rho({\bf r})$. The unconditional entropy of the macrostate  $\rho({\bf r})$  is defined  by the
Boltzmann formula
\begin{eqnarray}
S_0[\rho]=k_B\ln \Omega[\rho].
\label{h1}
\end{eqnarray}
The unconditional probability density of the density $\rho({\bf r})$ is therefore $P_0[\rho]\propto \Omega[\rho]\propto e^{S_0[\rho]/k_B}$. The number of complexions  $\Omega[\rho]$ can be obtained by a standard combinatorial analysis. For $N\gg 1$, when there is no microscopic constraints, using the Stirling formula, we find that the Boltzmann entropy is given by
\begin{eqnarray}
S_0[\rho]=-k_B\int \frac{\rho}{m}\ln\left (\frac{\rho}{Nm}\right )\, d{\bf r}.
\label{h2}
\end{eqnarray}
This is the same expression as in a perfect gas since the interaction between particles does not appear at that stage.

To evaluate the partition function (\ref{cano2}), instead of integrating over the microstates $\lbrace{\bf r}_1,...,{\bf r}_N\rbrace$, we can integrate over the macrostates $\rho({\bf r})$.  If we consider a system with long-range interactions so that a mean field approximation applies at the thermodynamic limit  $N\rightarrow +\infty$, introducing  the unconditional number of microstates $\Omega[\rho]$ corresponding to the macrostate $\rho$ [see Eqs. (\ref{h1}) and (\ref{h2})], and the potential energy $W[\rho]$ of the macrostate  $\rho$ [see Eq. (\ref{ybg12})], we obtain for $N\gg 1$:
\begin{eqnarray}
Z(\beta)\simeq e^{\frac{dN}{2}\ln \left (\frac{2\pi}{\beta m} \right )}\int e^{-\beta  W[\rho]}\Omega[\rho]\, \delta(M[\rho]-M)\, {\cal D}\rho\nonumber\\
 \simeq e^{\frac{dN}{2}\ln \left (\frac{2\pi}{\beta m} \right )}\int e^{S_0[\rho]/k_B-\beta W[\rho]}\, \delta(M[\rho]-M)\, {\cal D}\rho\nonumber\\
\simeq \int e^{-\beta F[\rho]} \, \delta(M[\rho]-M) \, {\cal D}\rho,\qquad
\label{h3}
\end{eqnarray}
where the free energy $F[\rho]$ is given by
\begin{eqnarray}
\label{h4}
F[\rho]=\frac{1}{2}\int\rho\Phi\, d{\bf r}+k_B T\int \frac{\rho}{m}\ln\left (\frac{\rho}{Nm}\right )\, d{\bf r}\nonumber\\
-\frac{d N}{2}k_B T\ln \left (\frac{2\pi k_B T}{m}\right ).
\end{eqnarray}
The canonical  probability density of the distribution $\rho$ is therefore
\begin{eqnarray}
P[\rho]=\frac{1}{Z(\beta)}e^{-\beta F[\rho]}\delta(M[\rho]-M).
\label{h5}
\end{eqnarray}

\subsection{The most probable macrostate}
\label{sec_mpm}

For systems with long-range interactions, for which  the mean field approximation is exact in the proper thermodynamic limit $N\rightarrow +\infty$, we have the extensive scaling $F\sim N$.  Accordingly, writing $F[\rho]=Nf[\rho]$ where $f\sim 1$, the partition function (\ref{h3}) may be  written as
\begin{eqnarray}
Z(\beta)= \int e^{-\beta N f[\rho]} \, \delta(M[\rho]-M) \, {\cal D}\rho.
\label{h6}
\end{eqnarray}
For $N\rightarrow +\infty$, we can make the saddle point approximation. We obtain
\begin{eqnarray}
Z(\beta)=e^{-\beta F(\beta)}\simeq e^{-\beta N f[\rho_*]},
\label{h7}
\end{eqnarray}
i.e.
\begin{eqnarray}
\lim_{N\rightarrow +\infty} \frac{1}{N}F(\beta)=f[\rho_*],
\label{h8}
\end{eqnarray}
where $\rho_*$ is the global minimum of free energy $F[\rho]$ at fixed mass. This corresponds to the most probable macrostate. We are led therefore to solving the minimization problem
\begin{eqnarray}
F(T)=\min_{\rho}\lbrace F[\rho]\, |\, M[\rho]=M  \rbrace.
\label{h9}
\end{eqnarray}

The previous results assume that there is a single global minimum of free energy at fixed mass. More generally, we shall be interested by possible local minima of free energy at fixed mass, which correspond to {\it metastable} states (the importance of these metastable states will be stressed in the sequel). The critical points of free energy at fixed mass are determined by the condition
\begin{eqnarray}
\label{h10}
\delta F+k_B T\alpha\delta M=0,
\end{eqnarray}
where $\alpha$ is a Lagrange multiplier associated with the conservation of
mass. Performing the variations, we obtain the mean field Boltzmann distribution
(\ref{ybg8}) where $A=Nme^{-\alpha m}$. A critical point of free energy 
at fixed mass is a (local) minimum if, and only if,
\begin{eqnarray}
\label{h11}
\delta^2 F=k_BT\int\frac{(\delta\rho)^2}{2\rho m}\, d{\bf r}
+\frac{1}{2}\int\delta\rho\delta\Phi\, d{\bf r}>0
\end{eqnarray}
for all perturbations $\delta\rho$ that conserve mass: $\delta
M=0$.

\subsection{The mean field Smoluchowski equation}
\label{sec_mfse}

We now derive kinetic equations governing the evolution of the average density
$\rho({\bf r},t)$ of the Brownian gas in interaction. From the $N$-body
Smoluchowski equation (\ref{smf3}), we can obtain the equivalent of the BBGKY
hierarchy \cite{hb2}:
\begin{eqnarray}
{\partial P_j\over\partial
t}=\sum_{i=1}^j\frac{\partial}{\partial {\bf r}_i}\biggl\lbrack D_*{\partial P_{j}\over\partial {\bf r}_{i}}+ \mu m^2 P_{j}\sum_{k=1,k\neq i}^{j} {\partial u_{i,k}\over\partial {\bf r}_{i}}\nonumber\\
+(N-j)\mu m^2\int P_{j+1}
{\partial u_{i,j+1}\over\partial {\bf r}_{i}}d{\bf r}_{j+1}\biggr\rbrack.
\label{h12a}
\end{eqnarray}
The stationary solutions of these equations coincide with the equations (\ref{ybg2}) of the YBG hierarchy. The first equation of this hierarchy is
\begin{eqnarray}
\label{h12b}
\frac{\partial P_1}{\partial t}=\frac{\partial}{\partial {\bf r}_1}\cdot\left\lbrack D_* \frac{\partial P_1}{\partial {\bf r}_1}+\mu m^2(N-1)\int \frac{\partial u_{12}}{\partial {\bf r}_1}P_2\, d{\bf r}_2\right\rbrack.\nonumber\\
\end{eqnarray}
This equation is exact but it is not closed.

For systems with long-range interactions,  we can neglect the correlations between the particles in the proper thermodynamic limit $N\rightarrow +\infty$ described previously \cite{cdr}. Therefore, the mean field approximation is exact and the $N$-body distribution function can be factorized, at any time t, in a product of $N$ one-body distribution functions:
\begin{eqnarray}
P_N({\bf r}_1,...,{\bf r}_N,t)=P_1({\bf r}_1,t)...P_1({\bf r}_N,t).\label{h12}
\end{eqnarray}
In particular,  $P_2({\bf r}_1,{\bf r}_2,t)=P_1({\bf r}_1,t)P_1({\bf r}_2,t)$. In that case, the first equation of the BBGKY hierarchy reduces to
\begin{eqnarray}
\label{h12c}
\frac{\partial P_1}{\partial t}&=&\frac{\partial}{\partial {\bf r}_1}\cdot\biggl\lbrack D_* \frac{\partial P_1}{\partial {\bf r}_1}\nonumber\\
&+&N\mu m^2 P_1({\bf r}_1,t)\int P_1({\bf r}_2,t)\frac{\partial u_{12}}{\partial {\bf r}_1}\, d{\bf r}_2\biggr\rbrack.
\end{eqnarray}
Therefore, the evolution of the average density $\rho({\bf r},t)=\langle
\sum_i m \delta({\bf r}-{\bf r}_i(t))\rangle=N m P_1({\bf r},t)$ is governed by
the  integro-differential equation
\begin{equation}
{\partial\rho\over\partial t}=D_*\Delta
\rho+\mu m \nabla\cdot\left\lbrack \rho\nabla\int u(|{\bf r}-{\bf r}'|)\rho({\bf
r}',t)\, d{\bf r}'\right\rbrack.
\label{h15}
\end{equation}
In can be rewritten as a mean field Smoluchowski equation \cite{hb2}:
\begin{eqnarray}
{\partial\rho\over\partial
t}=\nabla \cdot \left\lbrack {1\over\xi} \biggl
(\frac{k_B T}{m}\nabla\rho+\rho\nabla\Phi \biggr
)\right\rbrack, \label{h13}
\end{eqnarray}
where the self-consistent mean field potential $\Phi({\bf r},t)$ is given
by
\begin{eqnarray}
\Phi({\bf r},t)=\int u(|{\bf r}-{\bf r}'|)\rho({\bf r}',t)\, d{\bf r}'.
\label{h14}
\end{eqnarray}
The mean field Smoluchowski equation (\ref{h13}) satisfies an $H$-theorem for
the mean field free energy (\ref{h4}) which can be
obtained from Eq. (\ref{smf5}) by using the mean field approximation
(\ref{ybg5}). In terms of the free energy, the mean
field Smoluchowski equation (\ref{h13}) can be written as a gradient flow
\begin{eqnarray}
{\partial\rho\over\partial t}=\nabla\cdot \left\lbrack\frac{\rho}{\xi}\nabla
\left (\frac{\delta F}{\delta\rho}\right )\right\rbrack.
\label{h16}
\end{eqnarray}
A simple calculation gives
\begin{eqnarray}
\label{h17}
\dot F=-\int\frac{\rho}{\xi}\left\lbrack\nabla\left (\frac{\delta
F}{\partial\rho}\right )\right\rbrack^2\, d{\bf r}\nonumber\\
=-\int \frac{1}{\xi \rho}\left
(\frac{k_BT}{m}\nabla\rho+ \rho \nabla\Phi\right )^2\, d{\bf r}.
\end{eqnarray}
Therefore, $\dot F\le 0$ and $\dot F=0$ if, and only if, $\rho$ is the
mean field Boltzmann distribution (\ref{ybg8}) with the temperature of the bath $T$. Because of the
$H$-theorem, the system converges, for $t\rightarrow +\infty$, towards a
mean-field
Boltzmann distribution that is a (local) minimum of free energy
at fixed mass.\footnote{The steady states of
the mean field Smoluchowski equation are the critical points (minima,
maxima, saddle points) of the free energy $F[\rho]$ at fixed mass.
It can be shown \cite{nfp} that a critical
point of free energy is dynamically stable with respect to the mean
field Smoluchowski equation if, and only if, it is a (local)
minimum. Maxima are unstable
for all perturbations so they cannot be
reached by the system. Saddle points are unstable only for certain
perturbations so they can be reached if the system does not
spontaneously generate these dangerous perturbations.} If several minima exist
at the same temperature,
the selection depends on a notion of basin of attraction.

We note that the mean field free energy (\ref{h4}) may be written as
$F[\rho]=E[\rho]-T S[\rho]$ where
\begin{eqnarray}
\label{h4b}
E[\rho]=\frac{d}{2}Nk_B T+\frac{1}{2}\int\rho\Phi\, d{\bf r},
\end{eqnarray}
\begin{eqnarray}
\label{h4c}
S[\rho]=-k_B \int \frac{\rho}{m}\ln\left (\frac{\rho}{Nm}\right )\, d{\bf
r}\nonumber\\
+\frac{d}{2}Nk_B \ln\left (\frac{2\pi k_B T}{m}\right )+\frac{d}{2}Nk_B
\end{eqnarray}
are the energy and the entropy \cite{initialvalue}.

\subsection{The stochastic Smoluchowski equation}
\label{sec_sto}

In the preceding section we have considered the mean field limit $N\rightarrow
+\infty$ which amounts to neglecting the fluctuations. If the free energy
$F[\rho]$ has several minima (metastable states), and if $N$ is finite, the
system undergoes random transitions between the different metastable states due
to fluctuations. It explores the whole free energy landscape
and the distribution of the smooth (coarse-grained) density at equilibrium is
given by Eq. (\ref{h5}). It is therefore important to describe the fluctuations
(finite $N$ effects) giving rise to these random transitions.

Dean \cite{dean} has shown that the discrete density $\rho_d({\bf r},t)=\sum_i m \delta({\bf r}-{\bf r}_i(t))$ satisfies the stochastic equation
\begin{eqnarray}
\label{sto1}
\frac{\partial\rho_d}{\partial t}({\bf r},t)=D_*\Delta\rho_d({\bf r},t)\nonumber\\
+\mu m \nabla\cdot\left \lbrack\rho_d({\bf r},t)\nabla\int\rho_d({\bf r}',t)u(|{\bf r}-{\bf r}'|)\, d{\bf r}'\right \rbrack\nonumber\\
+\nabla\cdot \left \lbrack \sqrt{2D_*m\rho_d({\bf r},t)}\, {\bf R}({\bf
r},t)\right \rbrack,
\end{eqnarray}
where  ${\bf R}({\bf r},t)$ is a Gaussian white noise such that $\langle {\bf
R}({\bf r},t)\rangle={\bf 0}$ and $\langle
R^{\alpha}({\bf r},t)R^{\beta}({\bf r}',t')\rangle=\delta_{\alpha\beta}\delta({\bf r}-{\bf r}')
\delta(t-t')$. This equation is exact and bears the same information as the
$N$-body Langevin
equations (\ref{smf1}) or as the $N$-body Smoluchowski equation (\ref{smf3}). In
this sense, it contains
too much information. Furthermore, $\rho_d({\bf r},t)$  is a sum of Dirac
$\delta$-functions which is not easy to handle in practice.  If
we take the ensemble average of Eq. (\ref{sto1}) we obtain
\begin{equation}
\label{sto2}
\frac{\partial\rho}{\partial t}=D_*\Delta\rho+\mu m \nabla\cdot \int\langle\rho_d({\bf r},t)\rho_d({\bf r}',t)\rangle \nabla u(|{\bf r}-{\bf r}'|)\, d{\bf r}'.
\end{equation}
Using the identity
\begin{eqnarray}
\label{sto3}
\langle\rho_d({\bf r},t)\rho_d({\bf r}',t)\rangle=Nm^2P_1({\bf r},t)\delta({\bf r}-{\bf r}')\nonumber\\
+N(N-1)m^2P_2({\bf r},{\bf r}',t),
\end{eqnarray}
we see that this equation is equivalent to the first equation (\ref{h12b}) of
the BBGKY hierarchy. However, these equations are
not closed and they do not account for random
transitions between metastable states since they have been averaged.

In previous papers \cite{hb5,longshort}, we have argued that for systems
with long-range interactions, the evolution of the smooth
(coarse-grained) density $\rho({\bf r},t)$
is governed by the stochastic Smoluchowski equation
\begin{eqnarray}
\label{sto4}
\frac{\partial\rho}{\partial t}({\bf r},t)=D_*\Delta\rho({\bf r},t)\nonumber\\
+\mu m \nabla\cdot\left \lbrack \rho({\bf r},t)\nabla\int\rho({\bf r}',t)u(|{\bf r}-{\bf r}'|)\, d{\bf r}'\right \rbrack\nonumber\\
+\nabla\cdot \left \lbrack \sqrt{2D_*m\rho({\bf r},t)}\, {\bf R}({\bf
r},t)\right \rbrack.
\end{eqnarray}
This equation can be obtained from the theory of fluctuating hydrodynamics (see
Appendix B of \cite{hb5}). Although it has a similar mathematical form as Eq.
(\ref{sto1}), this equation is fundamentally  different from Eq. (\ref{sto1})
since it  applies to a {\it smooth} density $\rho({\bf r},t)$, not to
a sum of $\delta$-functions. It is also
different from Eqs. (\ref{h12b}) and (\ref{sto2}) since it is closed and not fully averaged. In a sense, it describes the evolution of the system at a mesoscopic level, intermediate between Eqs. (\ref{sto1}) and (\ref{sto2}) \cite{hb5,longshort}.

Introducing the mean potential (\ref{h14}), the stochastic Smoluchowski equation takes the form
\begin{eqnarray}
{\partial\rho\over\partial
t}=\nabla \cdot \left\lbrack {1\over\xi} \biggl
(\frac{k_B T}{m}\nabla\rho+\rho\nabla\Phi \biggr
)\right\rbrack+\nabla\cdot\left(\sqrt{\frac{2k_B T\rho}{\xi}}\, {\bf R}\right ).
\nonumber\\
\label{sto5}
\end{eqnarray}
In terms of the mean field free energy (\ref{h4}), it can  be rewritten as
\begin{eqnarray}
{\partial\rho\over\partial t}=\nabla\cdot \left\lbrack\frac{\rho}{\xi}\nabla
\left (\frac{\delta F}{\delta\rho}\right
)\right\rbrack+\nabla\cdot\left(\sqrt{\frac{2k_B T\rho}{\xi}}\, {\bf R}\right ).
\label{sto6}
\end{eqnarray}
Eq. (\ref{sto6}) may be interpreted as a stochastic Langevin equation for the field $\rho({\bf r},t)$. The corresponding Fokker-Planck equation for the probability density $P[\rho,t]$ of the density profile $\rho({\bf r},t)$ at time $t$ is
\begin{eqnarray}
\label{sto7}
&&\xi\frac{\partial P}{\partial t}[\rho,t]\nonumber\\
&=&-\int\frac{\delta}{\delta\rho({\bf r},t)}\left\lbrace \nabla\cdot \rho\nabla\left\lbrack {k_B T}\frac{\delta}{\delta\rho}+\frac{\delta F}{\delta\rho}\right\rbrack P[\rho,t]\right\rbrace\, d{\bf r}.\nonumber\\
\end{eqnarray}
Its stationary solution returns the canonical distribution (\ref{h5}) which shows the consistency of our approach.\footnote{The exact equation (\ref{sto1}) may be written in the form of Eqs. (\ref{sto5}) and (\ref{sto6}) where $\rho$ is replaced by $\rho_d$, $\Phi$ is replaced by $\Phi_d$, and $F$ is replaced by $F_d$. Accordingly, the evolution of $P[\rho_d,t]$ is given by a Fokker-Planck equation of the form of Eq. (\ref{sto7}) with the above-mentioned substitutions. It relaxes towards the distribution $P[\rho_d]=Z(\beta)^{-1} e^{-\beta F_d[\rho_d]}$ which is equivalent to the canonical distribution (\ref{cano1}). This result is valid for any $N$ and there is no need to use the Stirling formula (as emphasized by Dean \cite{dean}). For $N$ large, but not strictly infinite, the exact equation (\ref{sto1}) can be approximated by the stochastic Smoluchowski equation (\ref{sto4}) where now $\rho({\bf r},t)$ is a smooth density field. When $N\rightarrow +\infty$ we get the deterministic Smoluchowski equation (\ref{h13}) which ignores fluctuations. For finite $N$, the density probability of $\rho({\bf r},t)$ relaxes towards the equilibrium distribution $P[\rho]=Z(\beta)^{-1} e^{-\beta F[\rho]}$ which is valid for $N\gg 1$ and relies on the Stirling formula as discussed in Sec. \ref{sec_heur}. The relation between these different equations is further discussed in \cite{hb5,longshort}.} Actually, the form of the noise in Eq. (\ref{sto6}) may be determined precisely in order to recover the distribution (\ref{h5}) at equilibrium. We note that the noise is multiplicative since it depends on $\rho({\bf r},t)$ (it vanishes in regions devoid of particles).

Using a proper scaling as in Eq. (\ref{ybg4}), it can be shown that the noise term in Eq. (\ref{sto5})
is of order $1/\sqrt{N}$ so that it disappears in the $N\rightarrow +\infty$
limit. In that case, Eq. (\ref{sto5}) reduces to the mean field Smoluchowski
equation (\ref{h13}). If the free energy $F[\rho]$ has a single minimum, the
mean field Smoluchowski equation relaxes towards this minimum (this
corresponds to the  most probable macrostate). If the free energy $F[\rho]$ has
a several (local) minima, the mean field Smoluchowski equation relaxes
towards one of these minima and stays there for ever. However, when $N$ is
finite, we must take fluctuations into account and use the stochastic
Smoluchowski equation (\ref{sto5}). This equation describes random
transitions between the metastable states, establishing the canonical
distribution (\ref{h5}). In this paper, we solve the  stochastic
Smoluchowski equation (\ref{sto5}) for a system of self-gravitating Brownian
particles with a modified Poisson equation. This system displays a second order
phase transition below a critical temperature $T_c^*$ so that the free energy
$F[\rho]$ has a double-well structure and the particles undergo random
transitions between the two minima of this ``potential''. This
leads to a ``barrier crossing problem'' similar to the Kramers problem for the
diffusion of an overdamped particle  in a double well potential.

{\it Remark:} the stochastic Smoluchowski equation (\ref{sto6}) is different from the stochastic Ginzburg-Landau equation
\begin{eqnarray}
{\partial\rho\over\partial t}=-\Gamma\frac{\delta F}{\delta\rho}+\sqrt{2\Gamma k_B T}\zeta({\bf r},t),
\label{gl}
\end{eqnarray}
where $\zeta({\bf r},t)$ is a Gaussian white noise, used to describe the time-dependent fluctuations about equilibrium. Eq. (\ref{gl}) is a phenomenological equation because, in general, it is an impossible task to derive the true equation for the macroscopic variables directly from the dynamics of the microscopic variables of the system \cite{goldenfeld}. However, for Brownian particles with long-range interactions, this task is realizable and leads to the stochastic Smoluchowski equation (\ref{sto6}) instead of Eq. (\ref{gl}).

\subsection{The lifetime of metastable states}
\label{sec_lifetime}

The lifetime of a metastable state can be estimated by using an adaptation of
the Kramers formula \cite{risken}. Let us assume that below $T_c^*$ the free
energy $F[\rho]$ has two local minima $F_{meta}$  at the same height (metastable
states) separated by a saddle point $F_{saddle}$ (unstable). This is the case
for second order phase transitions. In order to pass from a metastable state to
the other the system has to cross a barrier of free energy played by the saddle
point. The
distribution of the smooth (coarse-grained) density $\rho({\bf r})$ at a fixed
temperature $T$ is given by Eq. (\ref{h5}). For a system initially prepared in a
metastable state, the probability for a fluctuation to drive it to a state with
density $\rho({\bf r})$ is $P[\rho]\sim e^{-\beta (F[\rho]-F_{meta})}$. If the
fluctuations bring the system in a configuration $\rho_{saddle}({\bf r})$
corresponding to the saddle point of free energy, it can then switch to the
other metastable state. Therefore, the lifetime of a metastable state may be
estimated by $t_{life}\sim 1/P[\rho_{saddle}]$, i.e. $t_{life}\sim
e^{\beta\Delta F}$ where $\Delta F=F_{saddle}-F_{meta}$ is the barrier of free
energy between the metastable state and the saddle point. For systems with
long-range interactions, in the proper thermodynamic limit $N\rightarrow
+\infty$, the free energy is proportional to $N$ so we can write $F[\rho]=N
f[\rho]$ where $f[\rho]\sim 1$. Therefore, we obtain the 
estimate\footnote{The formula (\ref{lifetime2})  was established
for self-gravitating Brownian particles in \cite{metastable} by generalizing the
Kramers approach \cite{kramers}. Actually, the derivation presented in
\cite{metastable} is not directly applicable to the present problem because it
corresponds to a small friction limit $\xi\rightarrow 0$ while we consider here
the case of a strong friction limit $\xi\rightarrow +\infty$. The formula
(\ref{lifetime2}) is valid in the two cases (only the prefactor is different)
because the qualitative arguments
presented above are very general. There are two possibilities to establish Eq.
(\ref{lifetime2}) in the present problem. One possibility is to use the Kramers
approach by starting directly from the Fokker-Planck equation (\ref{sto7}). This
will be considered in a future study. Another possibility, developed in Appendix
\ref{sec_instanton}, is to use the more modern instanton theory. This approach
makes clear that the Kramers formula is valid only in the weak noise limit (here
$N\gg 1$).}
\begin{eqnarray}
\label{lifetime2}
t_{life}\sim e^{N\beta\Delta f}.
\end{eqnarray}
Except in the vicinity of the critical point $T_c^*$ where $\Delta f\rightarrow
0$, the lifetime of a metastable state increases exponentially rapidly with the
number of particles, as $e^N$,  and becomes infinite in the thermodynamic limit
$N\rightarrow +\infty$ \cite{rieutord,art,metastable}. Therefore, for systems
with long-range interactions, metastable states have considerable lifetimes and
they can be regarded as stable states in practice. The probability to pass from
one metastable state to the other is a rare even since it scales as $e^{-N}$.
Random transitions between the two metastable states will be seen only close to
the critical point and/or for a sufficiently small number of particles (provided
that we wait long enough).

{\it Remark:} of course, these results can be generalized if the free energy has a local minimum (metastable state) and a global minimum of free energy (stable state) separated by a saddle point. This is the case for first order phase transitions. The lifetime of the metastable and stable states are still given by Eq. (\ref{lifetime2}). However, since the barrier of free energy between the saddle point and the stable state is larger than the barrier of free energy between the saddle point and the metastable state, the system will remain longer in the stable state than in the metastable state.

\section{Self-gravitating Brownian particles and bacterial populations with a modified Poisson equation}
\label{sec_formal}

\subsection{The Smoluchowski-Poisson system}

We consider a system of self-gravitating Brownian particles in a space of
dimension $d$. In the $N\rightarrow +\infty$ limit, we can ignore fluctuations.
In that case, the
evolution of the density of particles is governed by the Smoluchowski-Poisson
system
\begin{eqnarray}
\label{p1b}
\xi\frac{\partial\rho}{\partial t}=\nabla\cdot\left (\frac{k_B T}{m}\nabla\rho+\rho\nabla\Phi\right ),
\end{eqnarray}
\begin{eqnarray}
\label{p2b}
\Delta\Phi=S_d G \rho,
\end{eqnarray}
where $S_d$ is the surface of a unit sphere in $d$ dimensions. Its steady states are determined by the Boltzmann-Poisson equation
\begin{eqnarray}
\label{p3}
\Delta\Phi=S_d G A e^{-\beta m\Phi}.
\end{eqnarray}
These equations have been studied in \cite{sc}. When the system is enclosed within a spherical box of radius $R$ in order to prevent evaporation, the following results are found. In $d=3$, there is no global minimum of free energy. However, there exist a local minimum of free energy (metastable state) for $T>T_c$ where $T_c=GMm/(2.52Rk_B)$. For $T<T_c$ there is no critical point of free energy and the system undergoes an isothermal collapse leading to a Dirac peak. In $d=2$, there exist a global minimum of free energy for $T>T_c$ where $T_c=GMm/(4k_B)$.  For $T<T_c$ there is no critical point of free energy and the system undergoes an isothermal collapse leading to a Dirac peak. In $d=1$, there exist a global minimum of free energy for any $T\ge 0$. We note that the equilibrium states of Eqs. (\ref{p1b})-(\ref{p2b}) are spatially inhomogeneous. A homogeneous distribution is not a steady state of the ordinary SP system.

Following \cite{cd}, we consider a slightly different model of gravitational dynamics where the Poisson equation (\ref{p2b}) is replaced by the modified Poisson equation
\begin{eqnarray}
\label{p4}
\Delta\Phi=S_d G(\rho-\overline{\rho}),
\end{eqnarray}
where $\overline{\rho}=M/V$ is the mean density. The
steady states of the modified Smoluchowski-Poisson system defined by Eqs.
(\ref{p1b}) and (\ref{p4}) are determined by the modified Boltzmann-Poisson
equation
\begin{eqnarray}
\label{p5}
\Delta\Phi=S_d G \left (A e^{-\beta m\Phi}-\overline{\rho}\right ).
\end{eqnarray}
In that case, the uniform distribution $\rho=M/V$ and $\Phi=0$ is a steady state of the modified SP system for all temperatures. However, this uniform distribution is  stable only for $T>T_c^*(d)$, where $T_c^*(d)$ is a critical temperature depending on the dimension of space (see Sec. V of \cite{cd}). For $T<T_c^*(d)$ the system displays a second order phase transition from homogeneous to inhomogeneous states. In $d>1$, this second order phase transition adds to the ordinary isothermal collapse below $T_c$ mentioned above. In $d=1$, the modified SP system exhibits a second order phase transition while the usual SP system does not. These phase transitions have been discussed in detail in \cite{cd}.

If we take fluctuations (finite $N$ effects) into account, the evolution of the
density of particles is governed by the stochastic
Smoluchowski equation
\begin{eqnarray}
\label{p1}
\xi\frac{\partial\rho}{\partial t}=\nabla\cdot\left (\frac{k_B
T}{m}\nabla\rho+\rho\nabla\Phi\right )+\nabla\cdot\left(\sqrt{{2\xi k_B
T\rho}}\, {\bf R}\right )\quad
\end{eqnarray}
coupled to the Poisson equation (\ref{p2b}) or to the modified Poisson equation (\ref{p4}). For $N\rightarrow +\infty$, we can ignore the last term in Eq. (\ref{p1}) and we recover the deterministic Smoluchowski equation (\ref{p1b}). However, the stochastic Smoluchowski equation will be particularly relevant close to the critical points $T_c$ and $T_c^*$ where the mean field approximation is not valid due to the enhancement of fluctuations.

\subsection{The Keller-Segel model}
\label{sec_ks}

As discussed in \cite{cd}, the Smoluchowski-Poisson  system is connected to the Keller-Segel model describing the chemotaxis of bacterial populations in biology \cite{ks}. In general, the fluctuations are neglected leading to the deterministic Keller-Segel model
\begin{eqnarray}
\label{ks1}
\frac{\partial\rho}{\partial t}=\nabla\cdot\left (D\nabla\rho-\chi\rho\nabla c\right ),
\end{eqnarray}
\begin{eqnarray}
\label{ks2}
\frac{1}{D'}\frac{\partial c}{\partial t}=\Delta c-kc^2+\lambda\rho,
\end{eqnarray}
where $\rho({\bf r},t)$ is the density of bacteria and $c({\bf r},t)$ is the concentration of the secreted chemical (pheromone). The bacteria diffuse with a diffusion coefficient $D$ and they also experience a chemotactic drift with strength $\chi$ along the gradient of the chemical. The chemical is produced by the bacteria at a rate $D'\lambda$, is degraded at a rate $D'k^2$, and diffuses with a diffusion coefficient $D'$. In the limit of large diffusivity of the chemical ($D'\rightarrow +\infty$) and in the absence of degradation ($k=0$), the reaction-diffusion equation (\ref{ks2}) takes the form of a modified Poisson equation
\begin{eqnarray}
\label{ks3}
\Delta c=-\lambda(\rho-\overline{\rho}).
\end{eqnarray}
Therefore, this equation emerges naturally and rigorously in the biological problem in a well-defined limit. Furthermore, the ``box'' is justified in the biological problem since the bacteria are usually enclosed in a container. In that case, the Keller-Segel model (\ref{ks1}) and (\ref{ks3}) becomes isomorphic to the modified Smoluchowski-Poisson system (\ref{p1b}) and (\ref{p4}) provided that we make the correspondences
\begin{eqnarray}
\label{ks4}
\Phi=-c, \quad \chi=\frac{1}{\xi},\quad D=\frac{k_B T}{\xi m},\quad \lambda=S_d G.
\end{eqnarray}
In particular, the concentration $-c({\bf r},t)$ of the secreted chemical in
chemotaxis 
plays the same role as the gravitational potential $\Phi({\bf r},t)$ in
astrophysics. We can therefore map the SP system with a modified Poisson
equation into the KS model.

The ordinary KS model (\ref{ks1})-(\ref{ks2}) ignores fluctuations. In a
previous paper \cite{kssto}, we have proposed to
describe them with an equation of the form
\begin{eqnarray}
\label{ks5}
\frac{\partial\rho}{\partial t}=\nabla\cdot\left (D\nabla\rho-\chi\rho\nabla
c\right )+\nabla\cdot\left(\sqrt{{2D\rho m}}\, {\bf R}\right ),\quad
\end{eqnarray}
equivalent to the stochastic Smoluchowski equation (\ref{p1}).

{\it Remark:} Since the modified Poisson equation (\ref{ks3}) is justified in
biology, but 
not in astrophysics, it would seem more logical to present the following results
with the notations of biology. However, in order to make contact with previous
works on self-gravitating systems, it is preferable to use the notations of
astrophysics. These notations are also identical to those used in thermodynamics
and in the theory of Brownian motion while the notations used in biology for the
chemotactic problem are not directly related to thermodynamics.\footnote{The
interpretation of the Keller-Segel model in terms of thermodynamics is given in
\cite{nfp}.} Of course, our results can be immediately transposed to the problem
of chemotaxis by using the correspondences in Eq. (\ref{ks4}).

\section{The stochastic Smoluchowski-Poisson system}
\label{sec_ssp}

\subsection{The equation for the density}
\label{sec_efd}

We consider a gas of self-gravitating Brownian particles with the modified
Poisson equation (\ref{p4}). We assume that the system is enclosed within a
spherical box of radius $R$. The
evolution of the density $\rho({\bf r},t)$, taking fluctuations into
account, is described by the modified stochastic Smoluchowski-Poisson system
\begin{eqnarray}
\label{efd1}
\xi\frac{\partial\rho}{\partial t}=\nabla\cdot\left (\frac{k_B
T}{m}\nabla\rho+\rho\nabla\Phi\right )+\nabla\cdot\left(\sqrt{{2\xi k_B
T\rho}}\, {\bf R}\right ),\quad
\end{eqnarray}
\begin{eqnarray}
\label{efd2}
\Delta\Phi=S_d G(\rho-\overline{\rho}),
\end{eqnarray}
where $\overline{\rho}=M/V$ is the mean density. It is convenient to work with
dimensionless variables. As shown in Appendix \ref{sec_dimsp}, Eqs.
(\ref{efd1})-(\ref{efd2}) can be rewritten in dimensionless form as
\begin{eqnarray}
\label{efd3}
\frac{\partial\rho}{\partial t}=\nabla\cdot\left (
T\nabla\rho+\rho\nabla\Phi\right
)+\frac{1}{\sqrt{N}}\nabla\cdot\left(\sqrt{{2T\rho}}\, {\bf R}\right ),
\end{eqnarray}
\begin{eqnarray}
\label{efd4}
\Delta\Phi=S_d (\rho-\overline{\rho}),
\end{eqnarray}
where now $R=1$ and the density is normalized such that $\int \rho\, d{\bf
r}=1$. Therefore, $\overline{\rho}=d/S_d$. These equations can be obtained from
the original ones by setting $R=m=\xi=k_B=1$ and $G=1/N$, and by rescaling the
density by $N$ (i.e. we write $\rho=N\tilde{\rho}$ and finally note $\rho$
instead of $\tilde\rho$). This corresponds to the Kac scaling. In the
thermodynamic limit $N\rightarrow +\infty$, we have $E\sim N$, $T\sim 1$, $S\sim
N$ and $F\sim N$.\footnote{The proper thermodynamic limit of
self-gravitating systems corresponds to $N\rightarrow +\infty$ such that the
rescaled energy $\Lambda=-ER^{d-2}/GM^2$ and the rescaled temperature
$\eta=\beta GMm/R^{d-2}$
are of order unity \cite{cd}. This implies $S\sim E/T\sim F/T\sim N$. Different
scalings can then be considered (see
Appendix A of \cite{aakin}), the Kac scaling being the most convenient one.}

In this paper, we consider a one dimensional system and we make it symmetric with respect to the origin by imposing $R(-x,t)=-R(x,t)$. In that case, the foregoing equations reduce to
\begin{eqnarray}
\label{efd5}
\frac{\partial\rho}{\partial t}=\frac{\partial}{\partial x}\left ( T\frac{\partial\rho}{\partial x}+\rho\frac{\partial \Phi}{\partial x}\right )+\frac{1}{\sqrt{N}}\frac{\partial}{\partial x}(\sqrt{{2T\rho}}{R}),
\end{eqnarray}
\begin{eqnarray}
\label{efd6}
\frac{\partial^2\Phi}{\partial x^2}=2 (\rho-\overline{\rho}),
\end{eqnarray}
where $\overline{\rho}=1/2$. These equations can be solved in the domain $[0,1]$ with  the Neumann boundary conditions $\rho'(0,t)=\rho'(1,t)=0$ and $\Phi'(0,t)=\Phi'(1,t)=0$. The normalization condition is $2\int_0^1\rho(x,t)\, dx=1$.

\subsection{The equation for the mass profile}
\label{sec_efm}

We can reduce the two coupled equations (\ref{efd5}) and (\ref{efd6}) into a
single equation for the mass profile (or integrated density) defined by
\begin{eqnarray}
\label{efm1}
M(x,t)=2\int_0^x\rho(x',t)\, dx'.
\end{eqnarray}
We clearly have
\begin{eqnarray}
\label{efm2}
\frac{\partial M}{\partial x}=2\rho,\qquad \frac{\partial \Phi}{\partial x}=M(x,t)-x.
\end{eqnarray}
Integrating the stochastic Smoluchowski equation (\ref{efd5}) between $0$ and $x$, and using Eq. (\ref{efm2}), we find that the evolution of the mass profile is given by
\begin{eqnarray}
\label{efm3}
\frac{\partial M}{\partial t}=T\frac{\partial^2 M}{\partial x^2}+(M-x)\frac{\partial M}{\partial x}+2\sqrt{\frac{T}{N}\frac{\partial M}{\partial x}}{R}(x,t).
\end{eqnarray}
This stochastic partial differential equation (SPDE) has to be solved with the boundary conditions $M(0,t)=0$ and $M(1,t)=1$.

{\it Remark:} For the usual SP system where $M-x$ is replaced by $M$, we note that Eq. (\ref{efm3}) is similar to a noisy Burgers equation for a fluid with velocity $u(x,t)=-M(x,t)$ and viscosity $\nu=T$. This analogy is only valid in $d=1$.

\subsection{The equations of the $N$-body problem}
\label{sec_efn}

We may also directly solve the $N$-body dynamics. For the modified gravitational force in one dimension, the stochastic Langevin equations of motion of the $N$ Brownian particles are given by
\begin{eqnarray}
\frac{dx_i}{dt}=\frac{1}{\xi}F(x_i)+\sqrt{\frac{2k_B T}{m\xi}}R_i(t)
\label{efn1}
\end{eqnarray}
with
\begin{eqnarray}
\label{efn2}
F(x_i)=Gm\sum_{x_j\in S_+} {\rm sgn} (x_j-x_i)+G\overline{\rho}(2x_i-R).
\end{eqnarray}
We consider only particles in the interval $S_+=[0,1]$. The expression (\ref{efn2}) of the gravitational force is derived in Appendix \ref{sec_gm}. Using the scaling defined previously (see also Appendix \ref{sec_dimm}), these equations can be rewritten in dimensionless form as
\begin{eqnarray}
\label{efn3}
\frac{d x_i}{dt}=\frac{1}{N}\sum_j {\rm sgn} (x_j-x_i)+\frac{1}{2}(2x_i-1)+\sqrt{{2T}}R_i(t).\nonumber\\
\end{eqnarray}

\subsection{Second order phase transitions in $d=1$}
\label{sec_sec}

\begin{figure}
\begin{center}
\includegraphics[clip,scale=0.3]{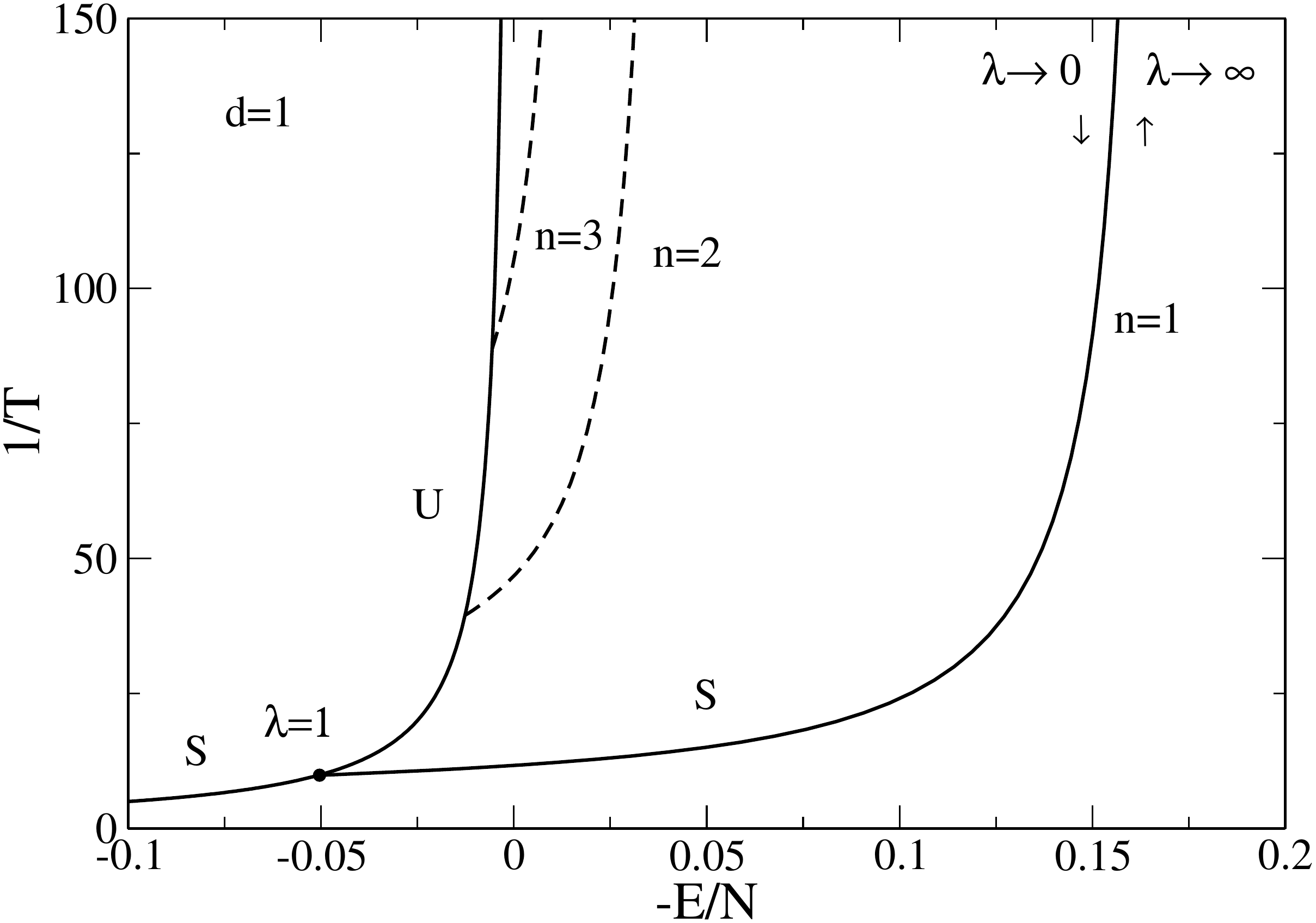}
\caption{Caloric curve giving the
energy per particle $-E/N$ as a function of the inverse temperature $1/T$. The
left branch corresponds to the homogeneous
phase. It is stable (S) for $T>T_c^*=1/\pi^2$ and unstable (U) for $T<T_c^*$. In
that case, it is replaced by an inhomogeneous phase which is stable. This
corresponds to the right branch denoted $n=1$. This branch is parameterized by
the density contrast $\lambda=\overline{\rho}/\rho_{0}$. It is degenerate in the
sense that, for a given temperature $T<T_c^*$, there exist two inhomogeneous
states with
the same energy but a different density contrast.}
\label{fig-d1-new}
\end{center}
\end{figure}

\begin{figure}
\begin{center}
\includegraphics[clip,scale=0.3]{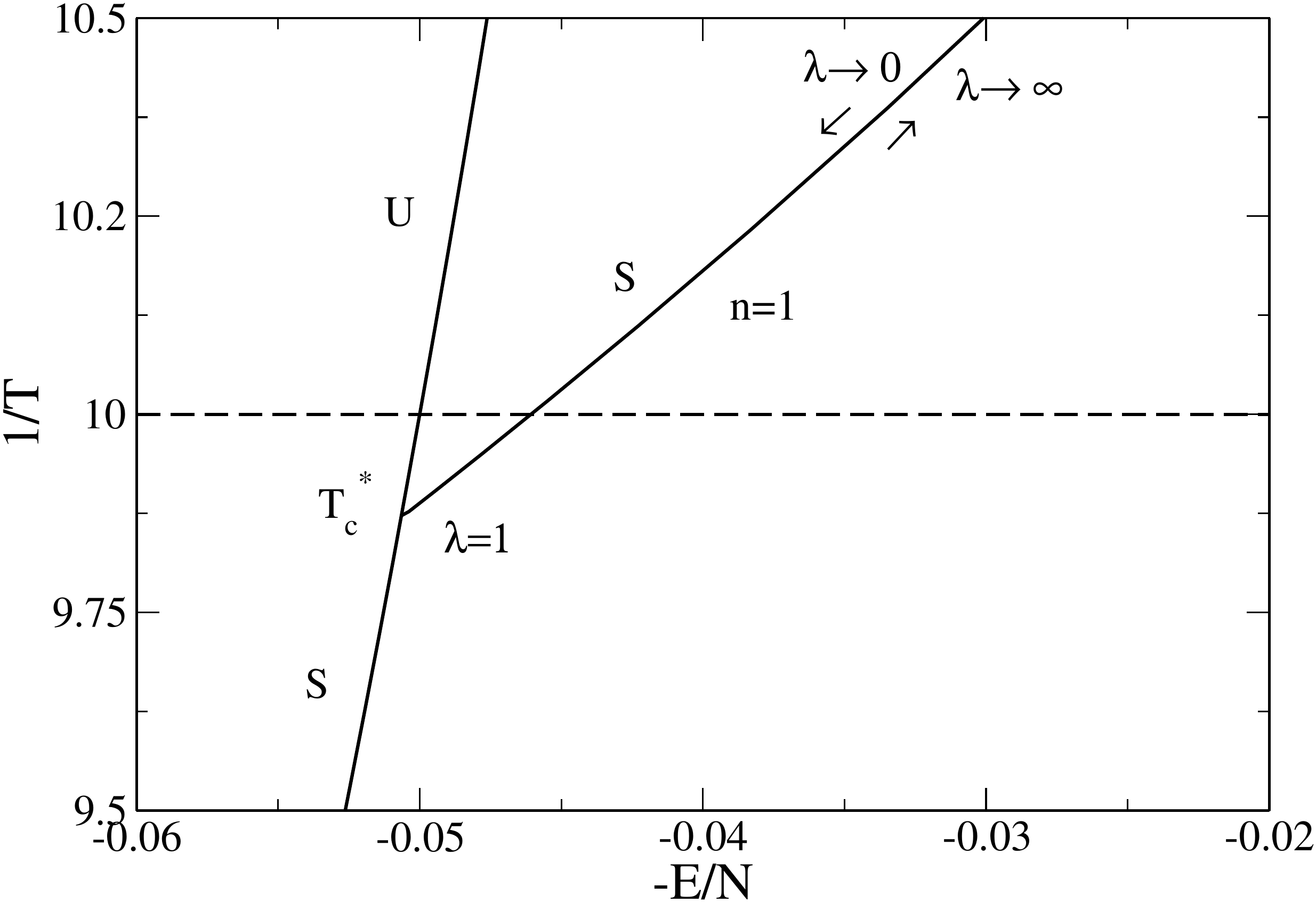}
\caption{Zoom of Fig. \ref{fig-d1-new} near the critical point
$T_c^*=1/\pi^2\simeq 0.101$. We have indicated by a dashed line the temperature
$T=0.1<T_c^*$ that will be considered in the section devoted to numerical
simulations.}
\label{eta-10}
\end{center}
\end{figure}

\begin{figure}[!h]
\begin{center}
\includegraphics[clip,scale=0.3]{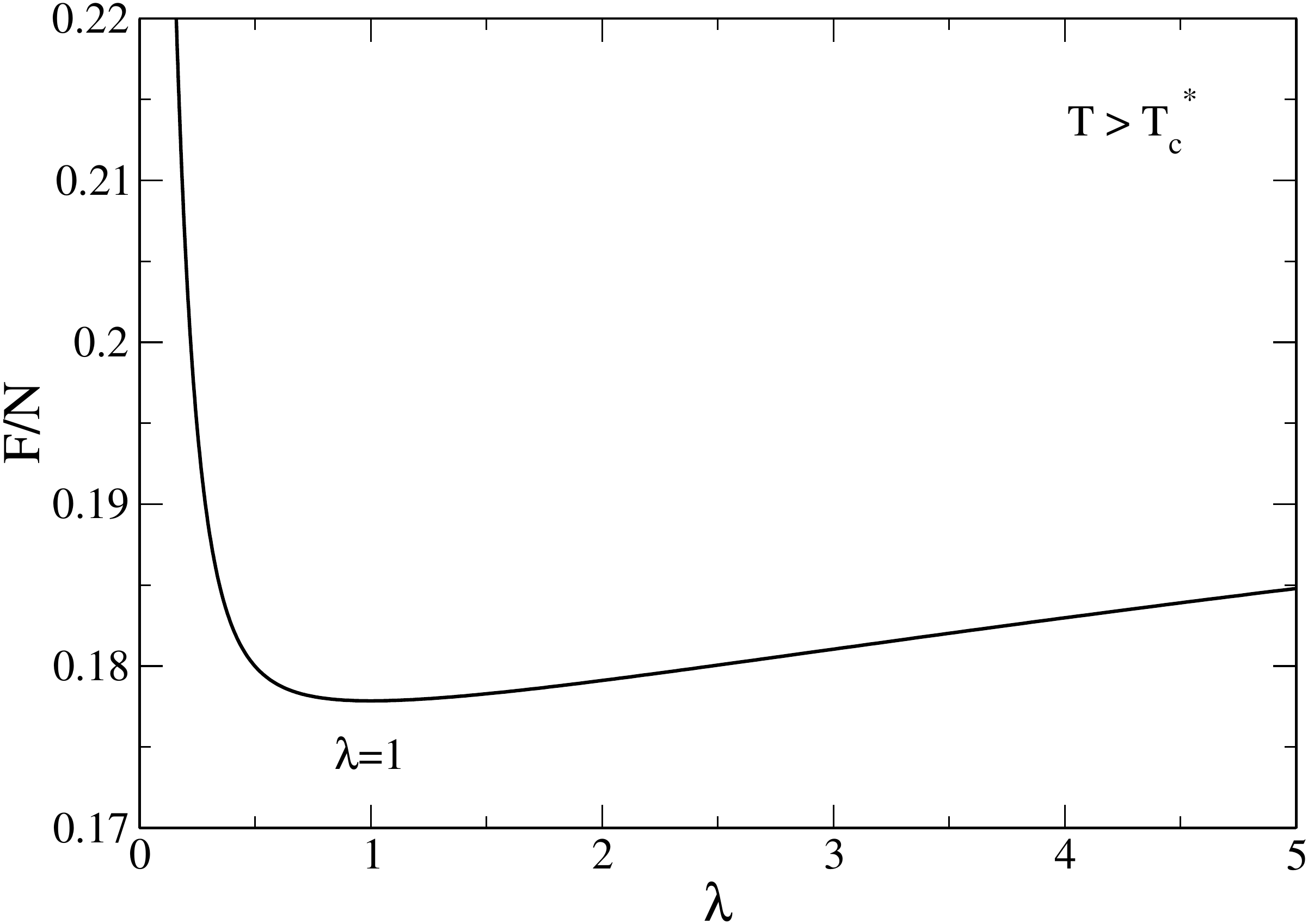}
\caption{Free energy $F(\lambda)=E(\lambda)-TS(\lambda)$ versus $\lambda$ for
$T=1/9\simeq 0.111>T_c^*$. It
presents a single minimum (stable state) corresponding to the homogeneous phase
($\lambda=1$). }
\label{free-ene-lambda-eta9}
\end{center}
\end{figure}

\begin{figure}[!h]
\begin{center}
\includegraphics[clip,scale=0.3]{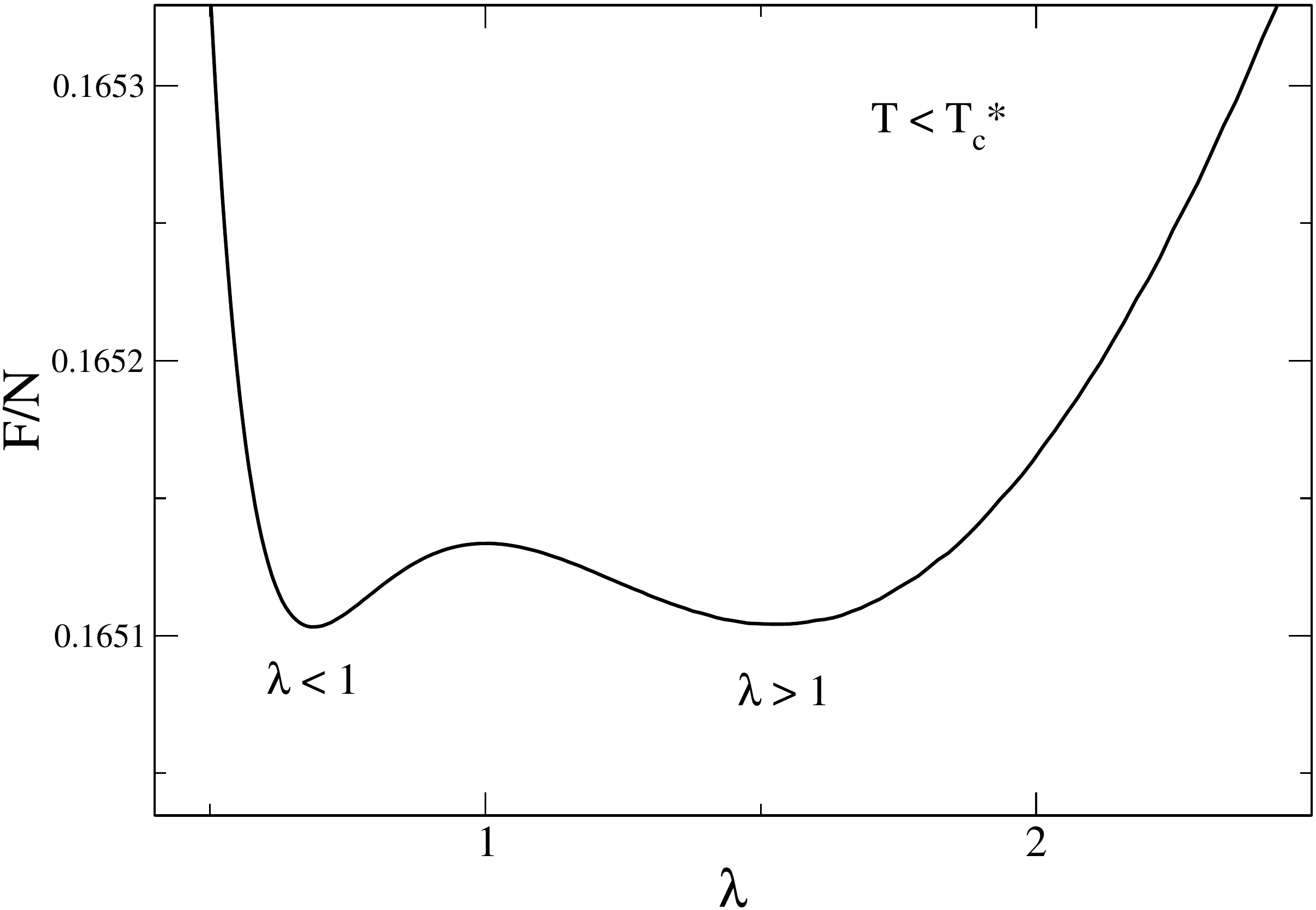}
\caption{Free energy $F(\lambda)=E(\lambda)-TS(\lambda)$  versus $\lambda$ for
$T=0.1<T_c^*$.
It has a double-well structure with two minima at
the same height (metastable states) and a local maximum (unstable state). The
local maximum corresponds to the homogeneous phase ($\lambda=1$). The minima
correspond to the inhomogeneous states concentrated near $x=0$ (specifically
$\lambda=0.69$) or near $x=1$ (specifically $\lambda=1.54$).
}
\label{free-ene-lambda-eta10}
\end{center}
\end{figure}

\begin{figure}
\begin{center}
\includegraphics[clip,scale=0.3]{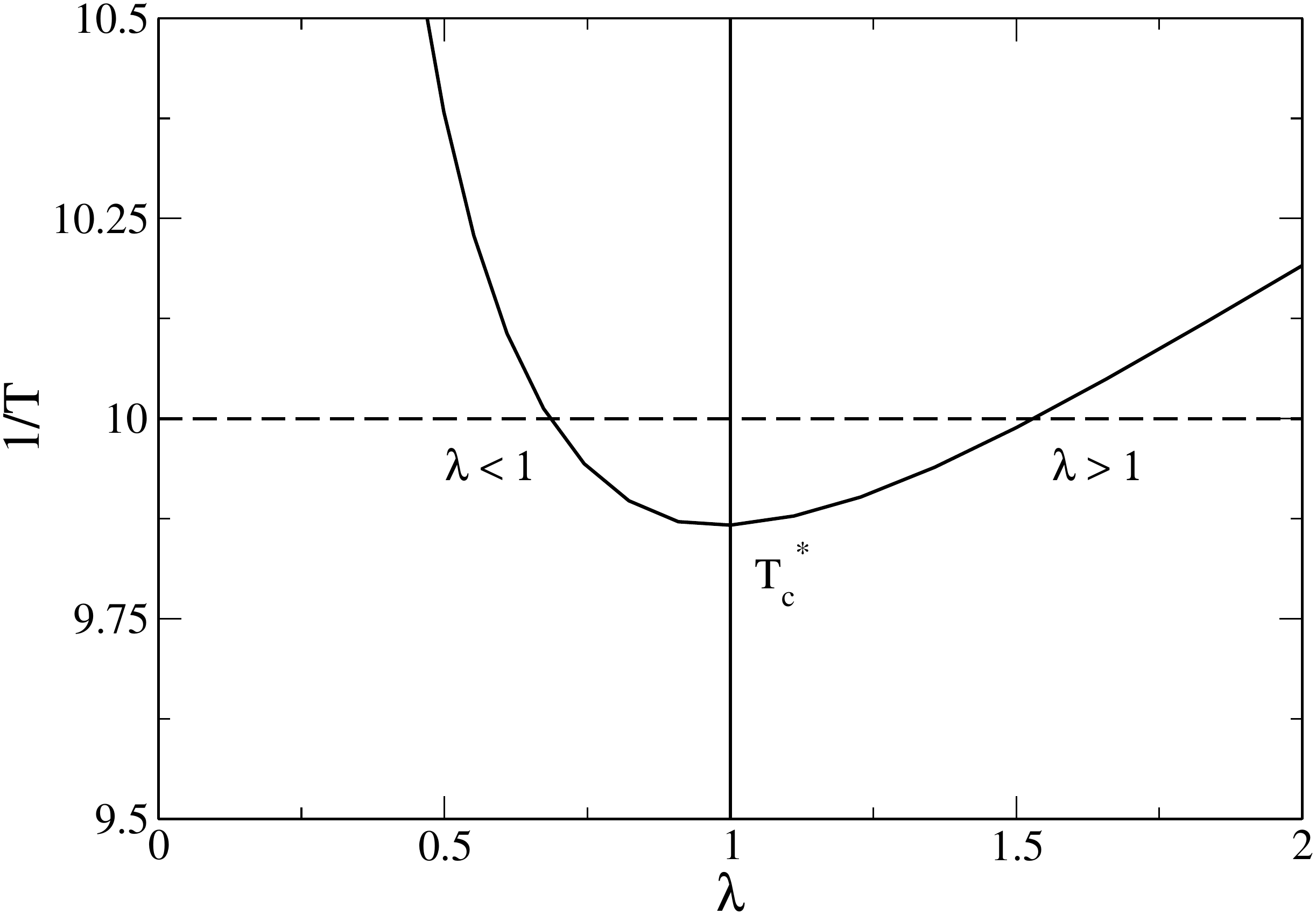}
\caption{Relation between the density contrast $\lambda$ and the inverse
temperature $1/T$.}
\label{etalambda}
\end{center}
\end{figure}

\begin{figure}
\begin{center}
\includegraphics[clip,scale=0.3]{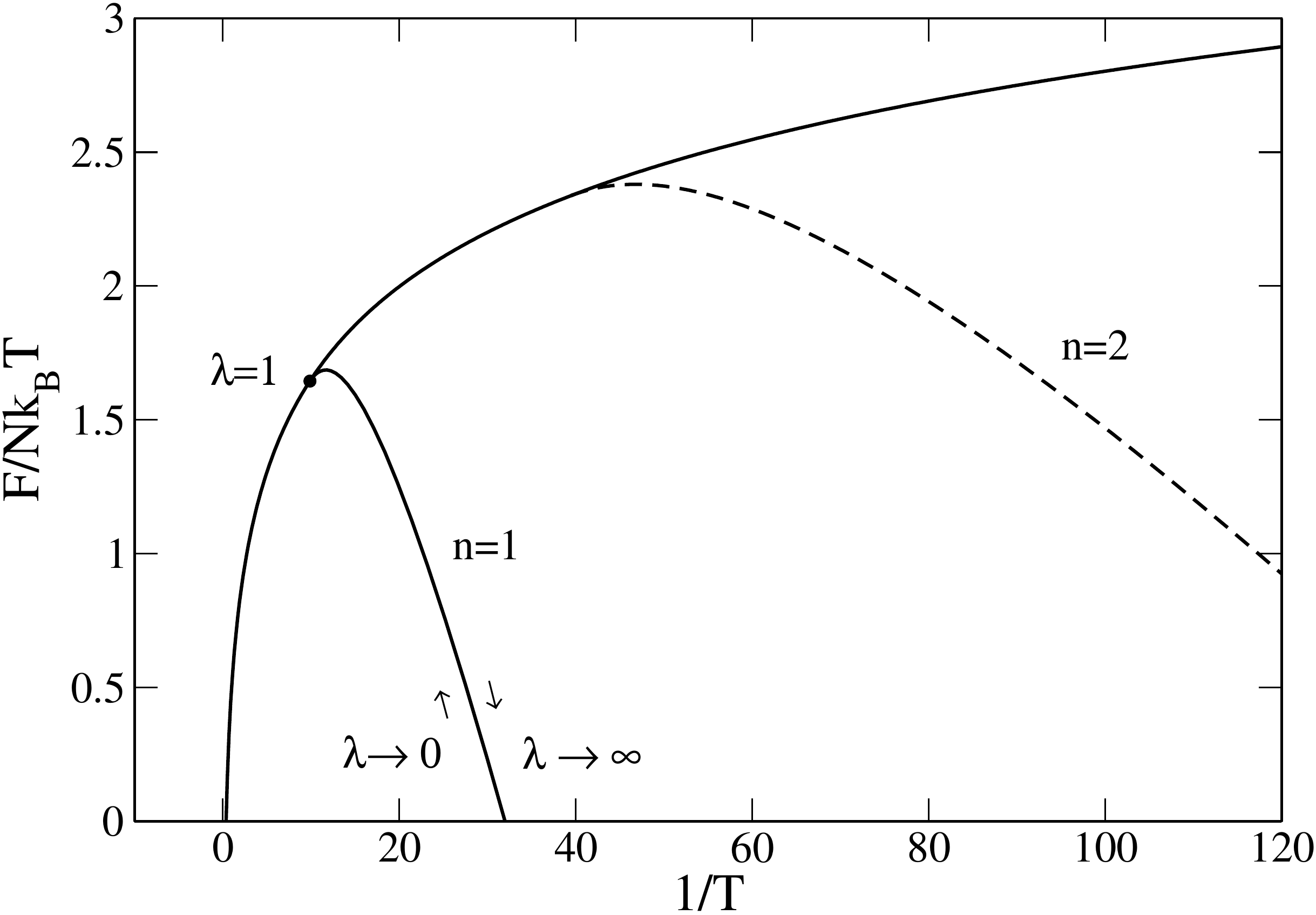}
\caption{Free energy $F$ as a function of the inverse temperature $1/T$.}
\label{free-d1}
\end{center}
\end{figure}

\begin{figure}[!h]
\begin{center}
\includegraphics[clip,scale=0.3]{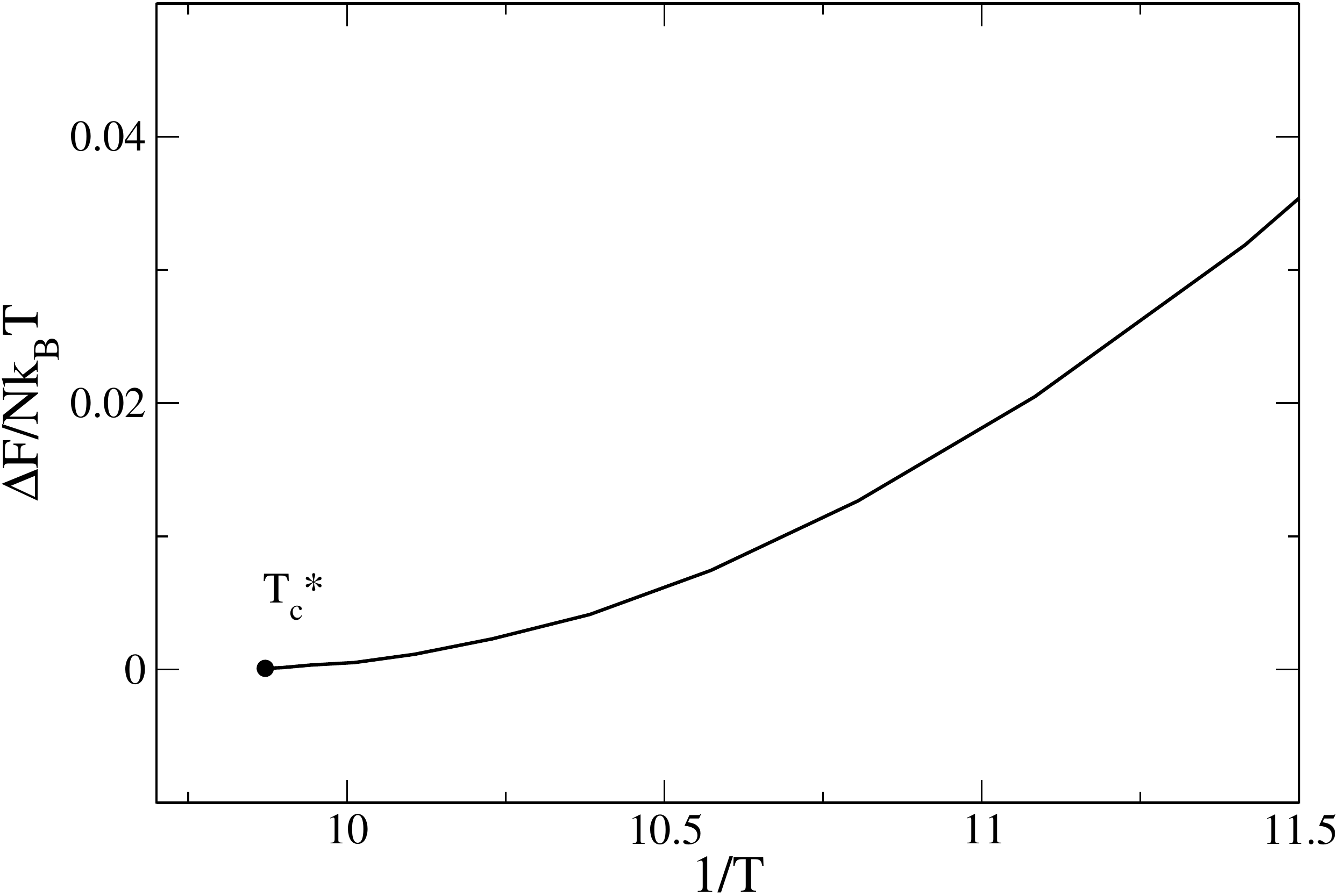}
\caption{Barrier of free energy $\Delta F/Nk_B T$, where $\Delta
F=F_{saddle}-F_{meta}$, as a
function of the inverse temperature $1/T$ close to the critical point.}
\label{deltafree-d1}
\end{center}
\end{figure}

The equilibrium states of the modified Smoluchowski-Poisson 
system [Eqs. (\ref{p1b}) and (\ref{p4})] are determined by the modified
Boltzmann-Poisson equation (\ref{p5}). They have been studied in detail in
\cite{cd}. In $d=1$, the caloric curve $E(T)$ giving the energy as a function of
the temperature is represented in Figs. \ref{fig-d1-new} and \ref{eta-10}. It
displays a second order phase transition at the critical temperature
$T_c^*=1/\pi^2$ marked by the discontinuity of $dE/dT$. For $T>T_c^*$ the system
is in the homogeneous phase which is the global minimum of free energy at fixed
mass (see Fig. \ref{free-ene-lambda-eta9}). For $T<T_c^*$ the homogeneous phase
is unstable and is replaced by an inhomogeneous phase corresponding to the
bifurcated branch $n=1$ in Figs. \ref{fig-d1-new} and \ref{eta-10} (as shown in
Fig. \ref{fig-d1-new}, there exist other bifurcated branches with $n$ clusters
but they are unstable \cite{cd}; only the branch with $n=1$ cluster that we are
considering is stable). Actually, this branch is degenerate. For $T<T_c^*$, the
free energy has two local minima (metastable states) at the same hight as shown
in Fig. \ref{free-ene-lambda-eta10}. They can be distinguished by the parameter
$\lambda=\overline{\rho}/\rho_0$ (density contrast) where $\rho_0=\rho(0)$ is
the central density. In the metastable state with $\lambda<1$, the particles are
concentrated near the center of the domain ($x=0$). In the metastable state with
$\lambda>1$, the particles are concentrated near the boundary of the domain
($x=1$). These inhomogeneous density profiles are shown in Fig.
\ref{profilesstochastique} for $T=0.1$. The density contrast  $\lambda$ of the
two metastable states is represented as a function of the temperature in Fig.
\ref{etalambda}. These two states have the same energy (see Figs. 
\ref{fig-d1-new} and \ref{eta-10}) and the same free energy (see Figs.
\ref{free-ene-lambda-eta10} and \ref{free-d1}).  The homogeneous phase
($\lambda=1$) corresponds to a saddle point of free energy (unstable). The
barrier of free energy $\Delta F/Nk_B T$ between the saddle point and the minima
is represented as a function of the temperature in Fig. \ref{deltafree-d1}.

For $N\rightarrow +\infty$, the evolution of the
density $\rho(x,t)$ is described by the  Smoluchowski equation
(\ref{p1b}) coupled to the modified Poisson equation (\ref{p4}). This
corresponds to the mean field approximation. For $T>T_c^*$, this equation
converges towards the homogeneous phase which is the unique minimum of free
energy at fixed mass (stable state). For $T<T_c^*$, this equation converges
towards one of the inhomogeneous states which is a local minimum of free energy
at fixed mass (metastable state). The choice of the metastable state depends on
the initial condition and on a notion of basin of attraction. For $N\rightarrow
+\infty$, the system remains in that state for ever. For ``small'' values of
$N$, or for $T$ close to the critical point $T_c^*$, the fluctuations become
important. For $T<T_c^*$ they induce random transitions between the two
metastable states discussed above. Such transitions can be described by the
modified stochastic Smoluchowski-Poisson system (\ref{efd5})-(\ref{efd6}) or,
equivalently, by the stochastic partial differential equation (\ref{efm3}) for
the mass profile.

\subsection{Numerical simulations of the modified stochastic SP system}
\label{sec_num}

We numerically solve Eq. (\ref{efm3}) using the finite differences method
described in \cite{miguel}. We work in the interval $[0,1]$. The boundary
conditions are $M(0,t)=0$ and $M(1,t)=1$. We start from a uniform distribution
$M(x,0)=x$.  We solve Eq. (\ref{efm3}) for $N=8000$, $10000$, and $12000$
particles. We choose the time step $\Delta t = 0.0001$ to make the integration
scheme stable, and we run each realization up to a time of order $10^{6}-10^{7}$
in our units. Performing this analysis for several values of $T$ and $N$ results
in very long simulations taking several months of CPU time. This is necessary to
have a good statistics and obtain clean results.

\begin{figure}
\begin{center}
\includegraphics[clip,scale=0.3]{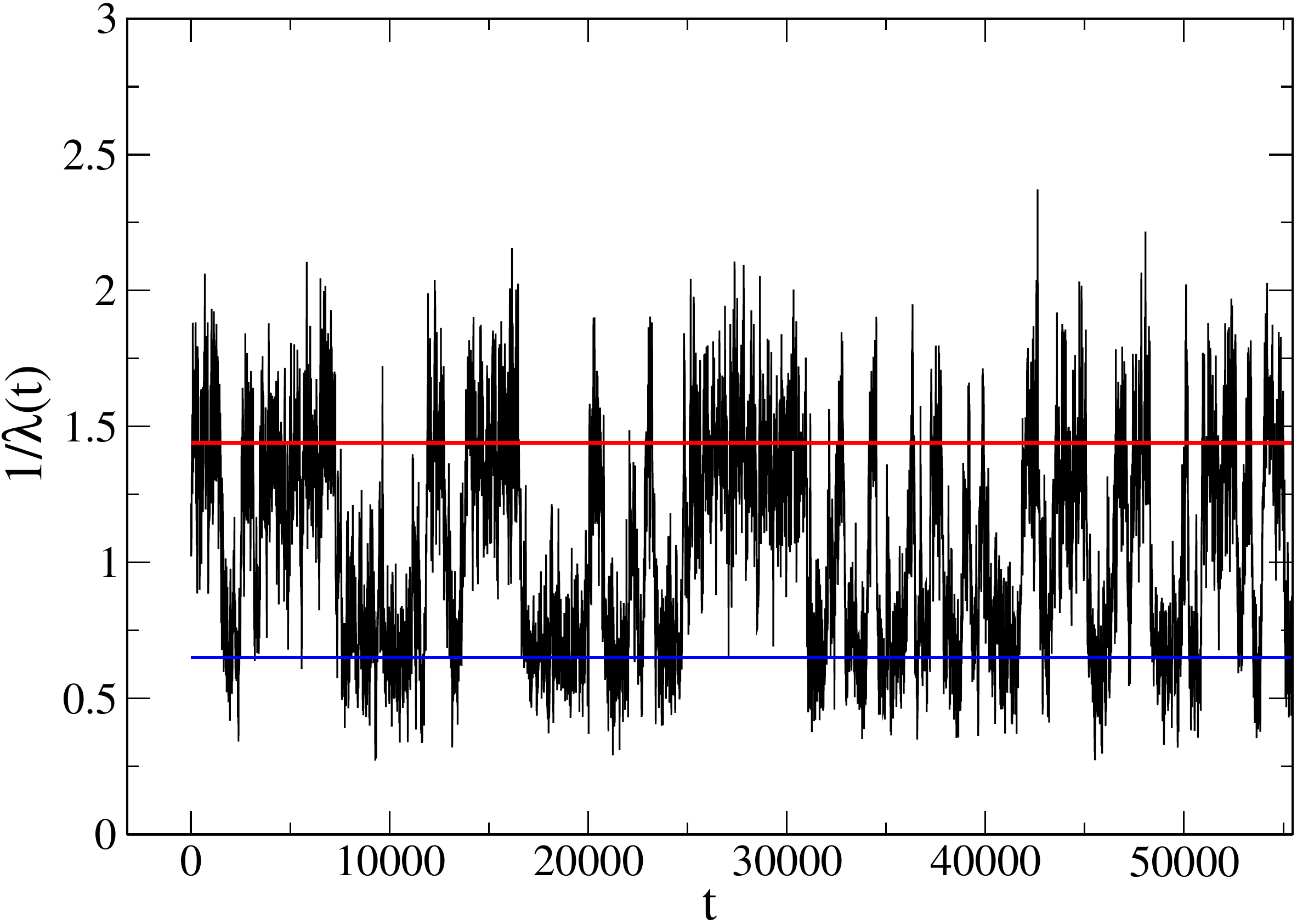}
\caption{Normalized central density $\rho(0,t)/\overline{\rho}=1/\lambda(t)$ as
a function of time for $T=0.1$ and $N=10000$. One observes clear signatures of
bistability. The system switches back and forth between two stable inhomogeneous
states around $\lambda=0.69$ (red) and $\lambda=1.54$ (blue).
The homogeneous phase is unstable.  Since the two stable configurations have the
same
value of free
energy, the average duration spent by the system in these two states is the
same.  }
\label{ro-eta-10}
\end{center}
\end{figure}

Fig. \ref{ro-eta-10} shows the temporal evolution of the normalized central
density $\rho(0,t)/\overline{\rho}=1/\lambda(t)$ for $T=0.1<T_c^*$ and
$N=10000$. The system undergoes random transitions between the metastable state
$\lambda=0.69$ and the metastable state $\lambda=1.54$. In the first case, the
density profile is concentrated around $x=0$ while in the second case it is
concentrated around $x=1$. The homogeneous phase ($\lambda=1$) is unstable.
Since the phase transition is second order, the two metastable states have the
same free energy and, consequently, the average time spent by the system in each
metastable state is the same. For $N\rightarrow +\infty$ the duration of the
plateau becomes infinite and the
system remains in only one of these states.

\begin{figure}[!ht]
\begin{center}
\includegraphics[clip,scale=0.7]{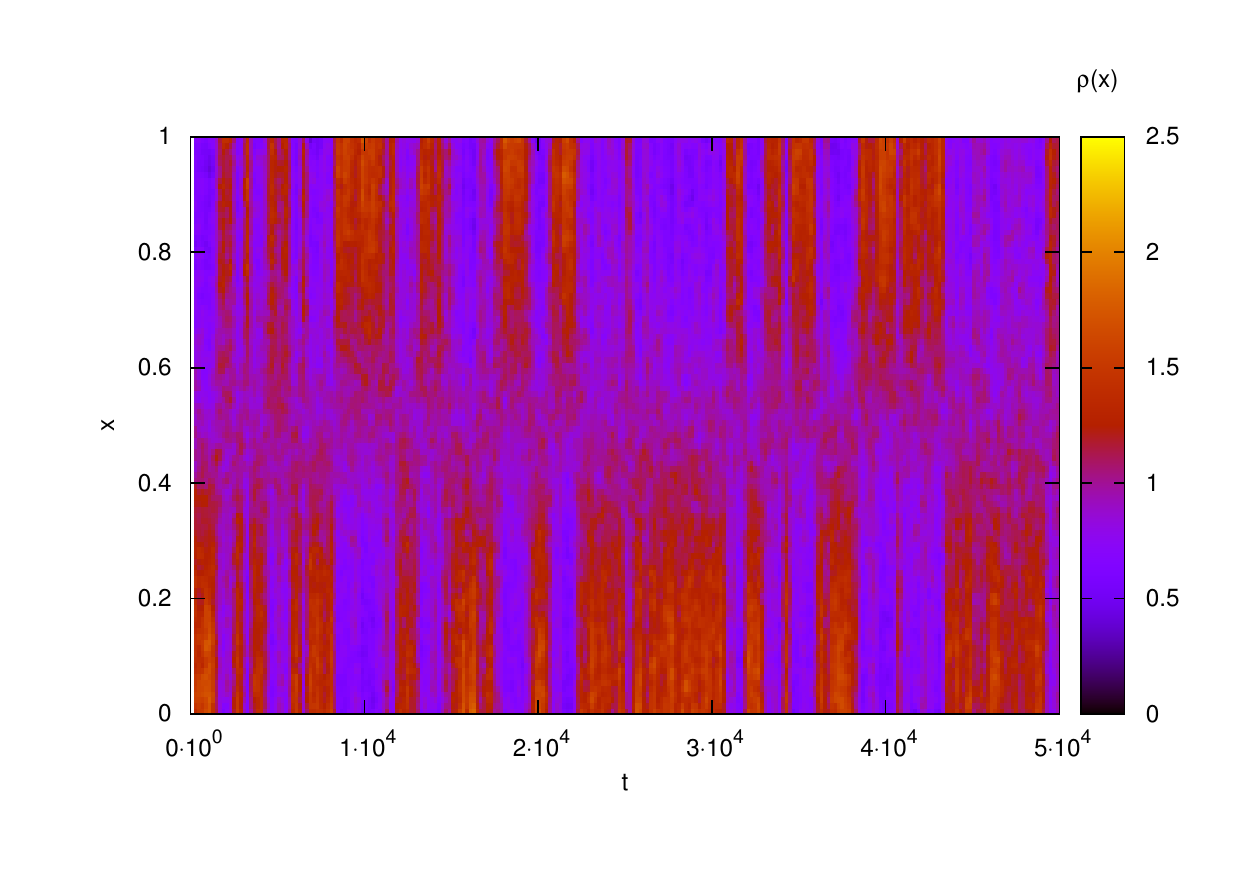}
\caption{Evolution of the density profile $\rho(x,t)$ as a function of time in a
spatio-temporal diagram $(x,t)$. High densities correspond to red
color and low densities to blue color.}
\label{dens-rid}
\end{center}
\end{figure}

Fig. \ref{dens-rid} shows the spatio-temporal evolution of the density 
profile $\rho(x,t)$. We clearly see the random displacement of the particles
from the center of the domain ($x=0$) to the boundary ($x=1$) and {\it vice
versa}.

\begin{figure}[!ht]
\begin{center}
\includegraphics[clip,scale=0.3]{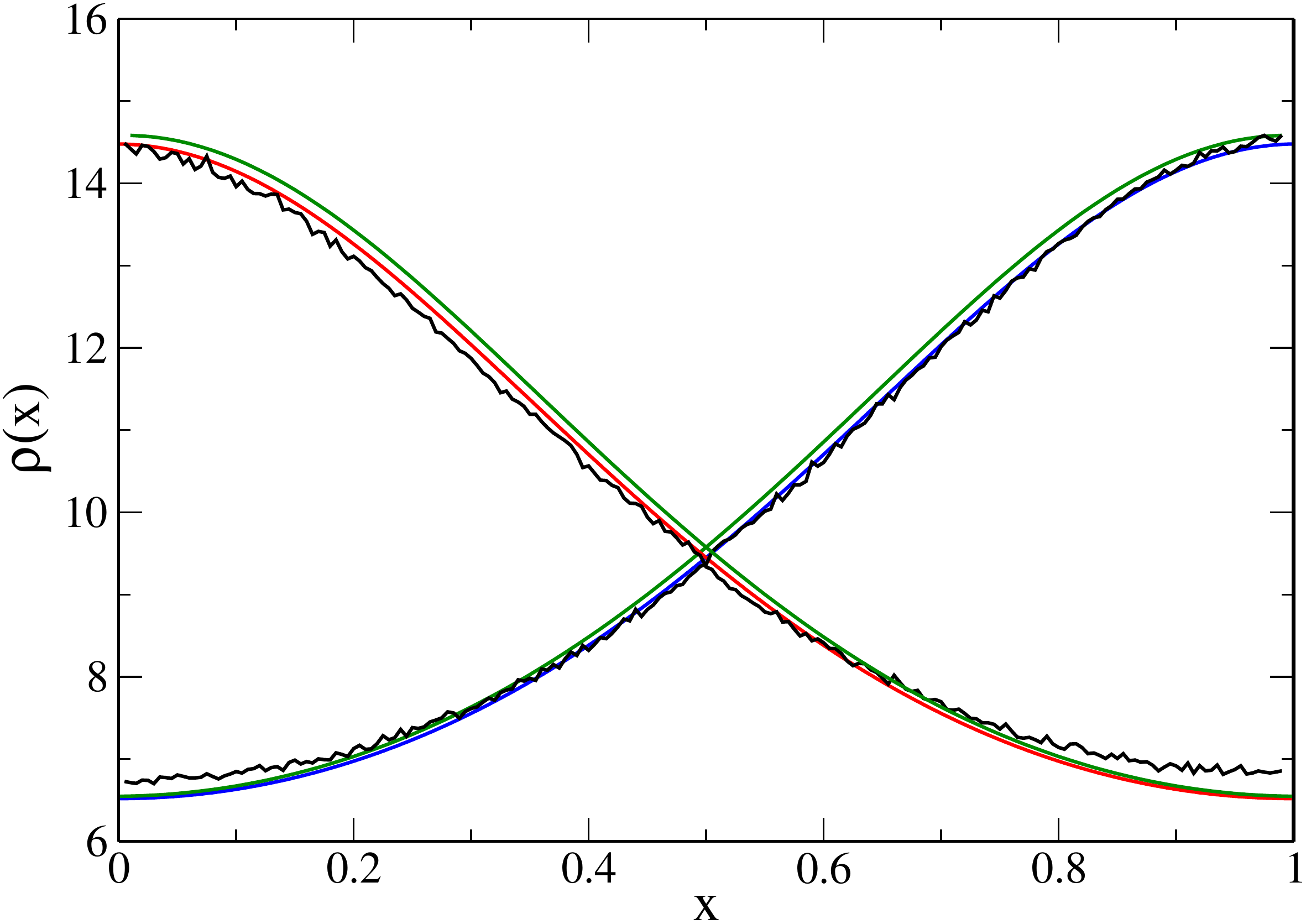}
\caption{Density profiles of the two metastable states for $T=0.1$ obtained by
solving the stochastic partial  differential equation (\ref{efm3}) with
$N=10000$ (green curves) or by solving the stochastic $N$-body equations (\ref{efn3}) with
$N=500$ (black curves). They are compared with the mean field equilibrium
distributions obtained in \cite{cd} for $N\rightarrow +\infty$ (red and blue curves).  }
\label{profilesstochastique}
\end{center}
\end{figure}

In Fig. \ref{profilesstochastique} we plot the density profiles of 
the two metastable states at $T=0.1$. The numerical profiles (green curves) are
obtained by solving the stochastic partial  differential equation (\ref{efm3})
with $N=10000$ and averaging the density in each phase ($\lambda<1$ or
$\lambda>1$) over an ensemble of very long simulations. They are compared with
the mean field profiles (red and blue curves) calculated in \cite{cd} showing a
very good agreement.  We have also plotted the numerical profiles (black curves)
obtained by directly solving the $N$-body equations (\ref{efn3}) with $N=500$.
The agreement is also good. The small differences are  probably due to finite
$N$ effects.

\begin{figure}
\begin{center}
\includegraphics[clip,scale=0.3]{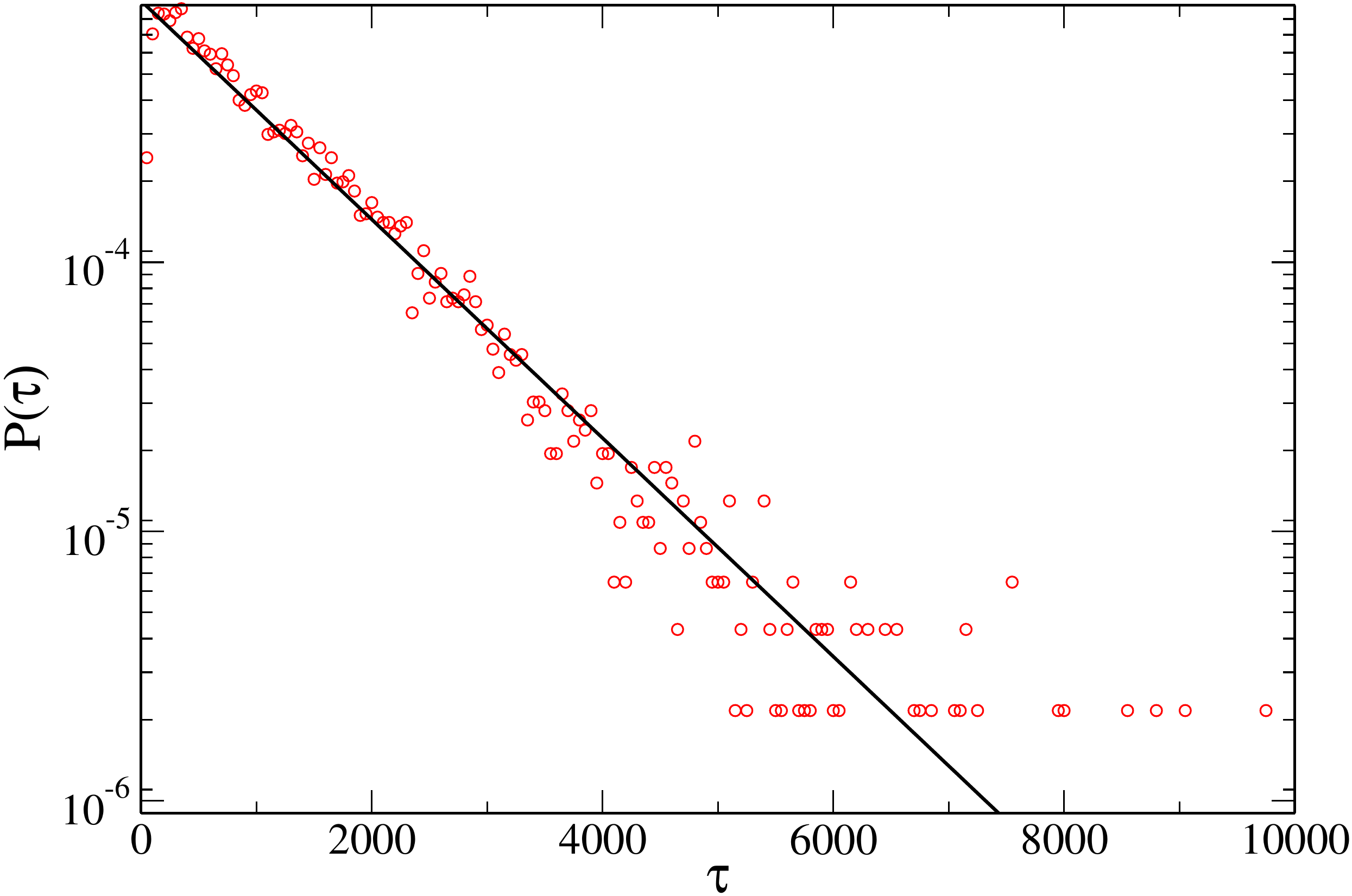}
\caption{Distribution of the residence time $\tau$ for $T=0.1$ and $N=10000$.
 The distribution is Poissonian indicating that the
random transitions from one metastable state to the other are statistically
independent.}
\label{isto-eta-10}
\end{center}
\end{figure}

In Fig. \ref{isto-eta-10} we have plotted the distribution of 
the residence time $\tau$ of the system in the metastable states. This is the
time spent by the system in a metastable state before switching to the other
($\tau$ is also called the ``first passage time'' from one state to the other).
If two successive transitions are statistically independent of one another, the
distribution of the residence time should be described by a Poisson process
\begin{eqnarray}
\label{num1}
P(\tau)=\frac{1}{\langle\tau\rangle} e^{-\tau/\langle\tau\rangle},
\end{eqnarray}
where $\langle \tau\rangle^{-1}$ gives the transition probability 
(per unit time) to switch from one state to the other. The average time spent by
the system in a metastable state is $\langle\tau\rangle$. The Poissonian
distribution of the residence time is confirmed by our numerical simulations. As
mentioned before, we need to run the simulations for a very long time
(especially for large $N$) in order to obtain a large number of transitions
between the two metastable states. Obtaining Fig. \ref{isto-eta-10} requires
several months of CPU time in order to obtain a good statistics for the
residence time.

\begin{figure}[!h]
\begin{center}
\includegraphics[clip,scale=0.3]{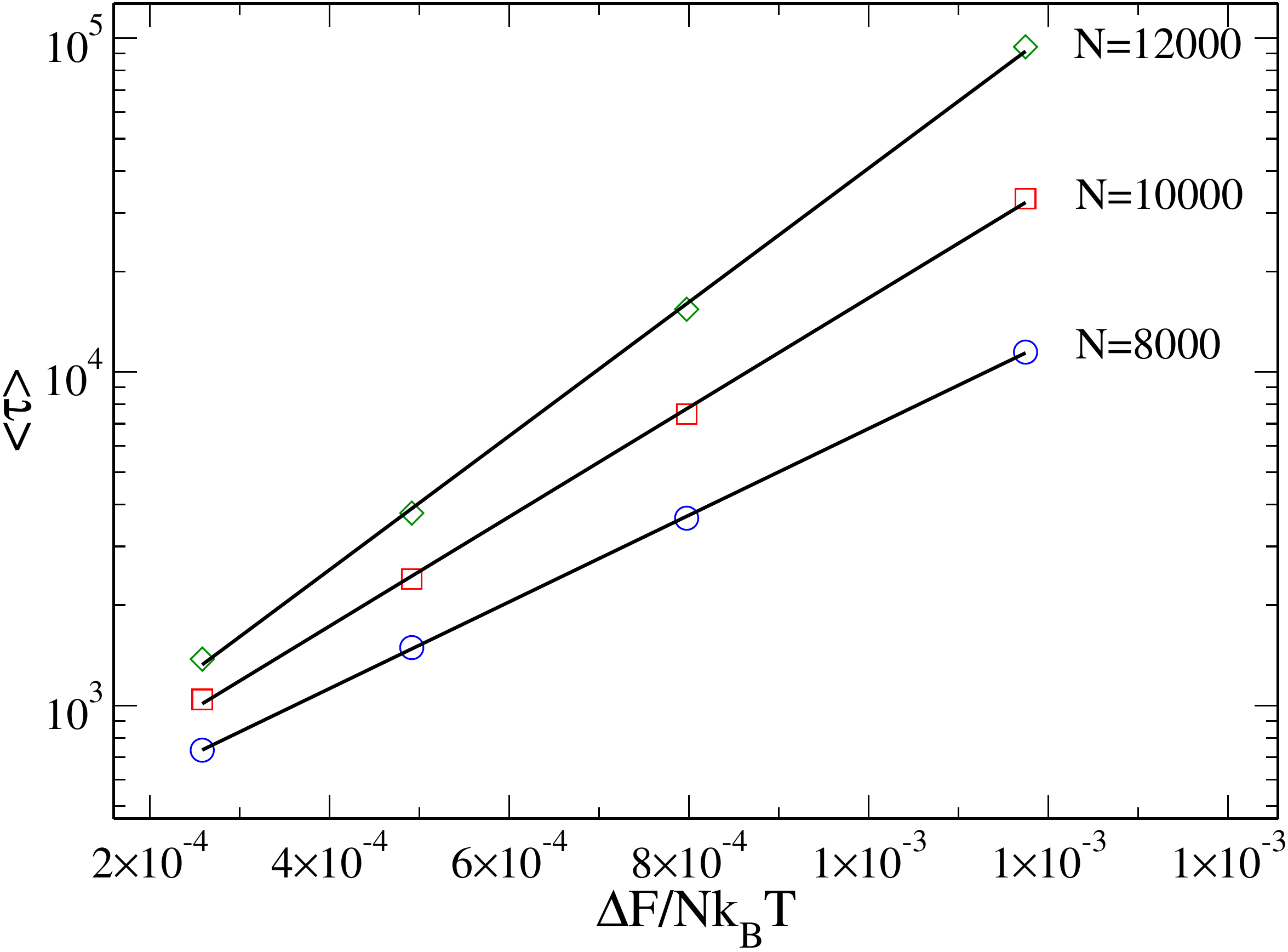}
\caption{Average time $\langle\tau\rangle$ spent by the system in a metastable state as a function of the barrier of free energy $\Delta f/k_B T$ for three different values of  $N$ in semi-log plot.}
\label{fig-tau}
\end{center}
\end{figure}

Since the system is at equilibrium with the thermal bath, we expect that the
average time $\langle\tau\rangle$ spent by the system in
a metastable state is given by the Arrhenius law
\begin{eqnarray}
\label{num2}
\langle\tau\rangle\propto e^{\Delta F/k_B T},
\end{eqnarray}
where $\Delta F=F_{saddle}-F_{meta}$ is the barrier of free energy between the
metastable state (inhomogeneous phase) and the unstable state (homogeneous phase).
This is similar to an activation process in chemical reactions. For a Brownian particle in a double-well potential
this formula has been established by Kramers \cite{kramers}. Here, the problem
is more complicated because the role of the particle $x(t)$ is played a field
$\rho(x,t)$ but the phenomenology remains the same. In the thermodynamic limit
$N\rightarrow
+\infty$, the barrier of free energy scales as $N$  and the dependence of the
barrier of free energy per particle $\Delta F/Nk_B T$ on the temperature  $T$
is represented in Fig. \ref{deltafree-d1}, using the results of \cite{cd}. To
test the law
\begin{eqnarray}
\label{num2b}
\langle\tau\rangle\propto e^{N\Delta f/k_B T},
\end{eqnarray}
we first fix $N$ and plot $\langle\tau\rangle$ as a function of $\Delta f/k_B T$ in semi-logarithmic coordinates by changing the temperature (see Fig. \ref{fig-tau}). We find a nice linear relationship confirming the exponential dependence of $\langle\tau\rangle$ with $\Delta f/k_B T$ for fixed $N$. According to this result, if we are close (resp. far) from the critical point $T_c^*$ the barrier of free energy is small (resp. large) and the system remains in a metastable state for a short (resp. long) time. This is essentially the meaning of the usual Kramers law.

\begin{figure}[!h]
\begin{center}
\includegraphics[clip,scale=0.3]{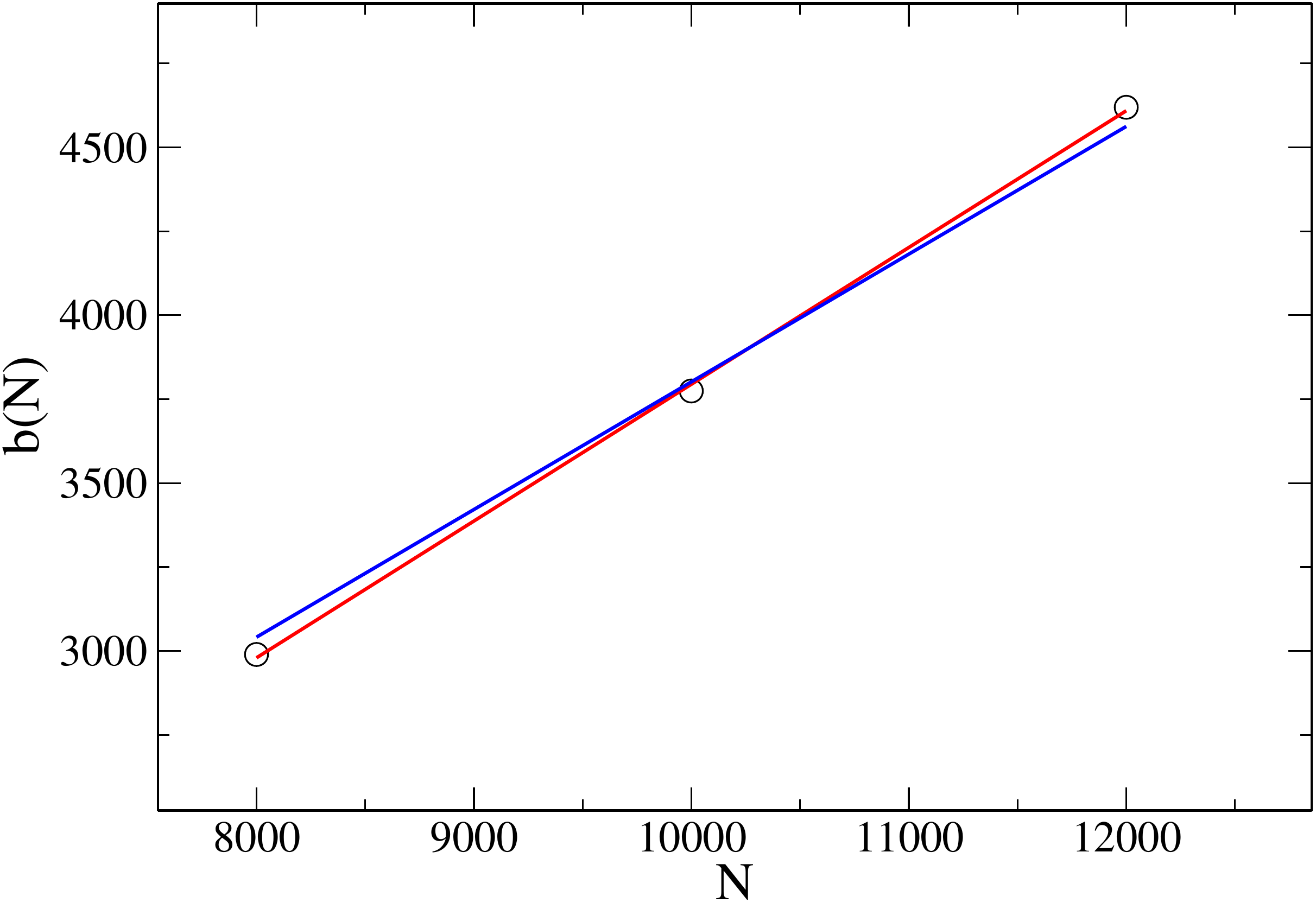}
\caption{Coefficient $b(N)$ as a function of $N$. We find $b(N)=0.38N$ (blue
line) or $b(N)=0.40 N-278$ (red line) depending on the fit.}
\label{BN}
\end{center}
\end{figure}

\begin{figure}[!h]
\begin{center}
\includegraphics[clip,scale=0.3]{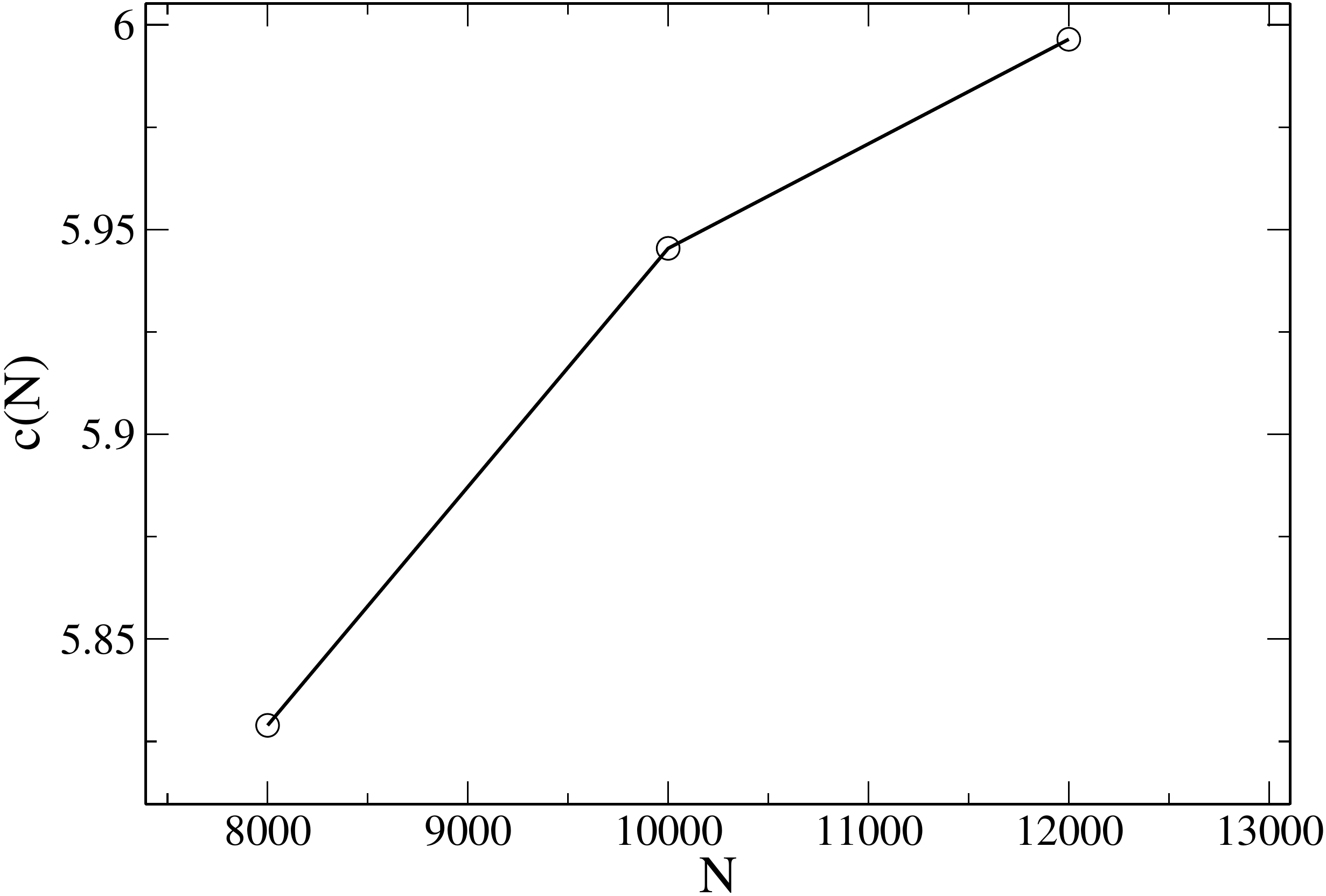}
\caption{Coefficient $c(N)$ as a function of $N$. It does not significantly vary
with $N$. This is compatible with a law $e^{c(N)}\propto
N^{\alpha}$ with $\alpha\in [0.3,0.5]$ but we clearly need more points to
be conclusive.}
\label{CN}
\end{center}
\end{figure}

We have also investigated the dependence of the 
coefficients with the number of particles $N$. To that purpose, we have written
Eq. (\ref{num2b}) in the more general form
\begin{eqnarray}
\label{num3}
\langle\tau\rangle=e^{b(N)\Delta f/k_B T+c(N)}.
\end{eqnarray}
The functions $b(N)$ and $c(N)$ obtained numerically are plotted in Figs.
\ref{BN} and \ref{CN}. We find that $b(N)$ varies linearly with $N$ while $c(N)$
does not change appreciably. We note, however, that $b(N)$ differs from the
relation $b(N)=N$ expected from Eq. (\ref{num2b}). We find $b(N)=0.38N$ or
$b(N)=0.40N-278$ depending on the fit. The reason of this difference with the
expected result $b(N)=N$ is not known. 
This may be a finite $N$ effect.\footnote{The validity of Eq.
(\ref{num2b}) requires $N\gg 1$. First, this is necessary in order to have
$\Delta F\sim N$. Actually, even if we impose this scaling for all $N$ (for
example by solving Eq. (\ref{efd3}) for any N although it is valid only for
$N\gg 1$) we still need $N\gg 1$ in order to justify the Kramers formula
(\ref{num2}). Indeed, as shown in Appendix \ref{sec_instanton}, this formula is
valid only in the weak noise limit which in our case corresponds to $N\gg 1$.
Therefore, $b(N)=N$ is valid only for $N\gg 1$ and $N=10000$ may not be large
enough. The observed relation $b(N)\simeq 0.4N$ may be a non-asymptotic result. 
It is also possible that the Kramers formula only gives $\ln
\langle\tau\rangle\propto N\Delta f/k_B T$ but does not determine the
coefficient because of non trivial prefactors.} It would be interesting to redo
the analysis for larger values of $N$ but this would require very long
simulations in order to achieve a good statistics.

According to the Arrhenius law (\ref{num2}) and the fact that the free energy scales as $N$ for systems with long-range interactions, we conclude that the lifetime of a metastable state scales as
\begin{eqnarray}
\label{num4}
\langle\tau\rangle\sim e^{N},
\end{eqnarray}
except close to a critical point where the barrier of 
free energy $\Delta f$ is small. As a result, for $N\gg 1$ (which is the norm
for systems with long-range interactions), metastable states have very long
lifetimes and, in practice, they can be considered as {\it stable} states. In
other words, metastable states (local minima of free energy) are as much, or
sometimes even more, relevant than fully stable states (global minima of free
energy) for systems with long-range interactions (see \cite{ijmpb,exclusion} for
an application of these considerations in astrophysics).  This exponential
dependence of the lifetime of the metastable states with the number of particles
for systems with long-range interactions \cite{rieutord,art,metastable} is a new
result as compared to the usual Kramers problem where there is only one
particle.

\subsection{Numerical simulations of the $N$-body equations}
\label{sec_numnbody}

\begin{figure}
\begin{center}
\includegraphics[clip,scale=0.3]{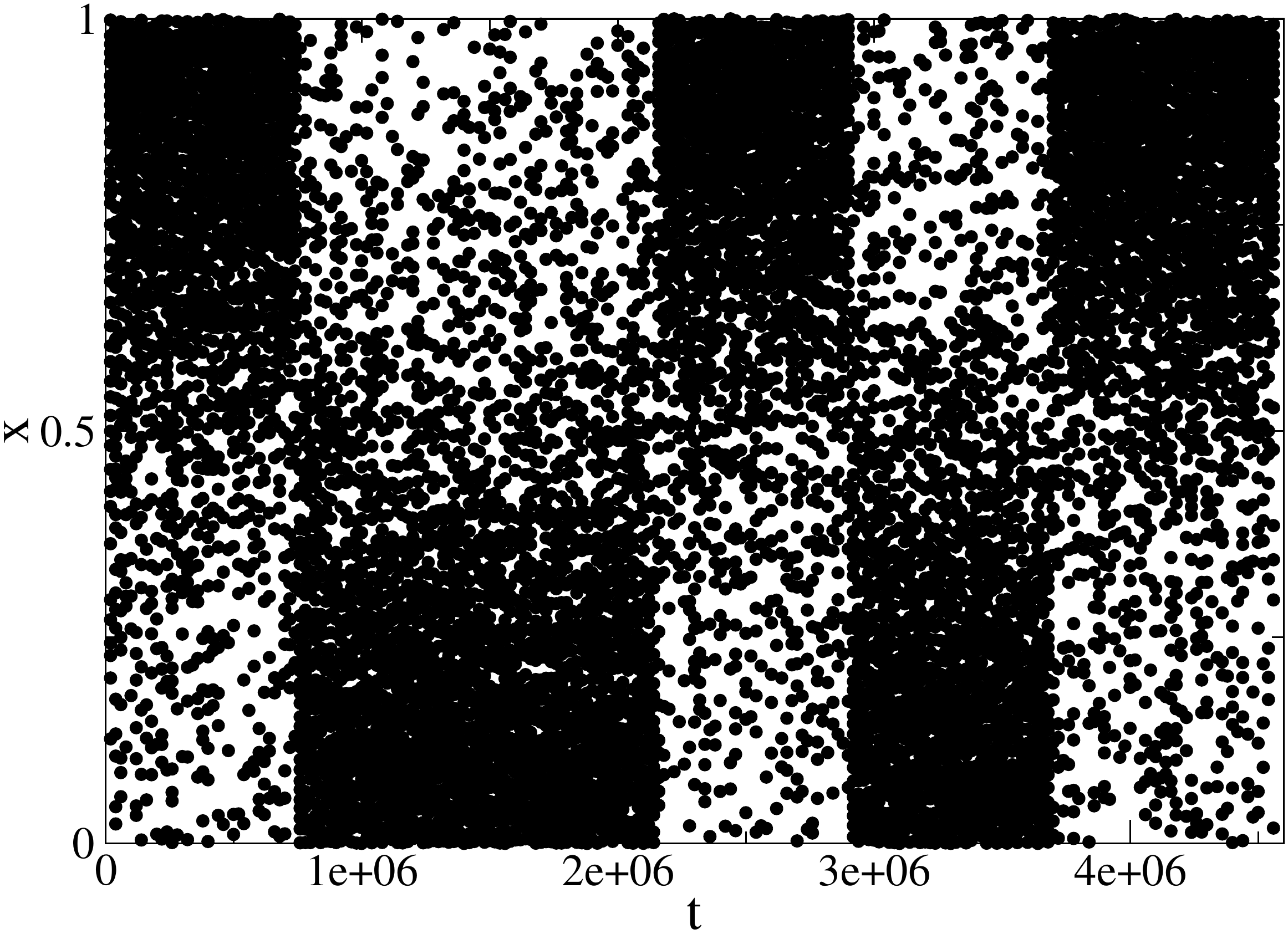}
\caption{Result of
N-body simulations for $T=1/12$ and $N=100$ particles. The particles switch back
and forth between the center of the domain ($x=0$) and the boundary ($x=1$).}
\label{y-eta12-rid}
\end{center}
\end{figure}

\begin{figure}
\begin{center}
\includegraphics[clip,scale=0.3]{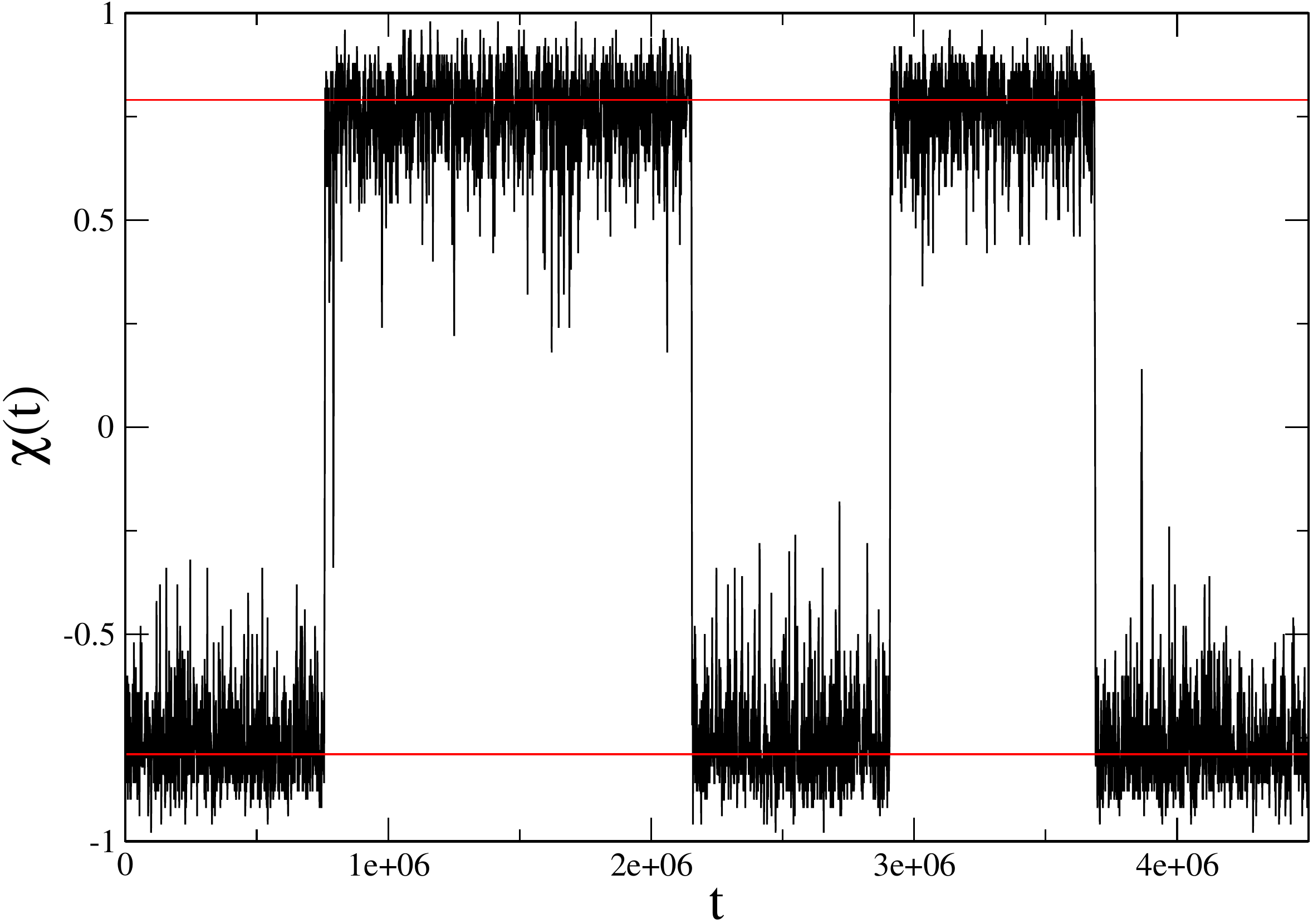}
\caption{Evolution of
the symmetrized mass $\chi(t)$ as a function of time
obtained by solving the stochastic $N$-body equations (\ref{efn3})  with $N=100$
and $1/T=12$.}
\label{chi-eta12-N100-rid}
\end{center}
\end{figure}

We have also performed direct numerical simulations of the stochastic $N$-body dynamics defined by Eq. (\ref{efn3}). Fig. \ref{y-eta12-rid} shows the spatio-temporal evolution of the particles for $T=1/12$ and $N=100$. We clearly see the random collective displacements of the particles between the center of the domain and the boundary. In Fig.  \ref{chi-eta12-N100-rid} we plot the time evolution of the symmetrized mass $\chi(t)=2\int_{0}^{1/2}\rho_d(x,t)\, dx-1$. By definition, $\chi=1$ if all the particles are in the interval $[0,1/2]$ and $\chi=-1$  if all the particles are in the interval $[1/2,1]$. The symmetrized mass can be used as an order parameter similar to the density contrast in Fig. \ref{ro-eta-10}.

\section{Conclusion}

We have studied a one dimensional model of self-gravitating Brownian particles
with a modified Poisson equation \cite{cd}. This model also describes the
chemotaxis of bacterial populations in the limit of high diffusivity of the
chemical and in the absence of degradation. It presents a second order phase
transition from a homogeneous phase to an inhomogeneous phase below a critical
temperature $T_c^*$. For $T<T_c^*$, it displays a bistable behavior. The
particles switch back and forth between the center of the domain and the
boundary. These two configurations correspond to metastable states that are
local minima of the mean field free energy $F[\rho]$ at
fixed mass. This leads to a ``barrier crossing problem'' similar
to the Kramers problem for the diffusion of an overdamped particle  in a double
well potential. We have shown with theoretical arguments and numerical
simulations that the phenomenology of these random transitions is similar to
that of the usual Kramers problem. In particular, the mean lifetime of the
system in a metastable state is proportional to the exponential of the barrier
of free energy $\Delta F$ divided by  $k_B T$. However, a specificity of systems
with long-range interactions (with respect to the Kramers problem) is that the
barrier of free energy is proportional to $N$ so that the typical lifetime of
a metastable state is considerable since it scales as $e^N$ (except close to a
critical point). Therefore, metastable states are stable states in practice
\cite{rieutord,art,metastable}. The probability to pass from one metastable state to the other is a rare even since it scales as $e^{-N}$. Random transitions between metastable states
can be seen only close to the critical point, for moderate
values of $N$, and for sufficiently long times.
This is the situation investigated in this paper.

The present results have been obtained for a particular model but they are expected to be valid for any system with long-range interactions presenting second order phase transitions. Another example is the Brownian mean field (BMF) model \cite{bmf}. Below a critical temperature $T_c=1/2$ it undergoes random transitions  between two states with magnetization $M_x=1$ and $M_x=-1$ if we enforce a symmetry with respect to the $x$-axis (i.e. if we impose $M_y=0$). Similar results should also be obtained for systems with long-range interactions presenting first order phase transitions. In that case, the system exhibits random transitions between a stable state (global minimum of free energy) and a metastable state (local minimum of free energy). The switches between these two states are asymmetric and the system remains longer in the stable state that corresponds to the global minimum of free energy. Apart from that asymmetry, the phenomenology is similar to the one reported in this paper. An example of systems with long-range interactions presenting first order phase transitions is provided by a 3D self-gravitating gas in a box with a small-scale exclusion constraint \cite{ijmpb,exclusion}.

In the present work, and in the other examples quoted above, the system 
is at equilibrium with a thermal bath. The case of systems with long-range
interactions that are maintained out-of-equilibrium by an external forcing is
also interesting \cite{nardini,campakuramoto}. These systems present a similar
phenomenology except that the equivalent of the external potential $V(x)$ in the
usual Kramers problem, or the mean field free energy $F[\rho]$ in the present
problem, is not obvious at first sight and requires a specific treatment.

A domain of physical interest where random transitions between
different attractors occur concerns two-dimensional fluid flows \cite{bv}.
Different types of phase transitions have been evidenced in that context. For
example, phase transitions in a flow with zero circulation and low energy
enclosed in a rectangular domain have been investigated in \cite{jfm96}. In a
domain of aspect ratio $\tau<\tau_c=1.12$, there exist degenerate metastable
states having the form of ``monopoles''  in which positive vorticity is
concentrated in the center of the domain and negative vorticity is distributed
in the periphery, or the opposite. These two symmetric metastable
states are separated by a saddle point that has the form of a ``dipole''. In a
domain of aspect ratio $\tau>\tau_c$ there exist degenerate metastable states
having the form of ``dipoles'' in which positive vorticity is located on the
left of the domain and negative vorticity is located on the right, or the
opposite. These two symmetric metastable states are separated by a saddle point
having the form of a ``monopole''. If the flow is modeled as a gas of point
vortices, it will exhibit random transitions between the two metastable states
due to finite $N$ effects. These random transitions will be particularly
important close to the critical point $\tau_c$ where the barrier of entropy
between the metastable states and the saddle point is low. This is very similar
to the present problem. Random transitions may also be observed between two
attractors if the system is stochastically forced at small scales. For
example, random
transitions between a dipole and a jet have been observed in \cite{bs} for 2D
fluid flows described by the stochastic Navier-Stokes equation in a periodic
domain. These random transitions are related to those studied in the present
paper but the problem is physically different because, in that case, the system
is out-of-equilibrium and there is no obvious form of thermodynamical
potential.

\appendix

\section{The dimensionless stochastic Smoluchowski-Poisson system}
\label{sec_dimsp}

The dimensional stochastic Smoluchowski-Poisson system is given by Eqs. (\ref{efd1})-(\ref{efd2}). It conserves the total mass $M=Nm=\int\rho\, d{\bf r}$.
If we introduce dimensionless variables $n$, $x$, $\phi$, $\eta=1/\theta$, $\tau$ and $Q$ through the relations
\begin{eqnarray}
\label{dimsp1}
\rho=\frac{M}{R^d}n,\quad {\bf r}=R{\bf x},\quad \Phi=\frac{GM}{R^{d-2}}\phi,\nonumber\\
 \eta=\frac{\beta GMm}{R^{d-2}}=\frac{1}{\theta},\quad t=\frac{\xi R^d}{GM}\tau,\quad {\bf R}=\sqrt{\frac{GM}{\xi R^{2d}}}{\bf Q},
\end{eqnarray}
and recall that the volume of a $d$-dimensional sphere of radius $R$ is $V=\frac{1}{d}S_d R^d$, we obtain
\begin{eqnarray}
\label{dimsp2}
\frac{\partial n}{\partial \tau}=\nabla\cdot\left ( \theta\nabla n+n\nabla\phi\right )+\frac{1}{\sqrt{N}}\nabla\cdot\left(\sqrt{{2\theta n}}{\bf Q}\right ),
\end{eqnarray}
\begin{eqnarray}
\label{dimsp3}
\Delta\phi=S_d (n-\overline{n}),
\end{eqnarray}
where $\overline{n}=d/S_d$ is the mean density of a sphere of unit mass and
unit radius. The conservation of mass is expressed by $\int n\, d{\bf x}=1$. On
the other hand, ${\bf Q}({\bf x},\tau)$ is a Gaussian white noise satisfying
$\langle {\bf Q}({\bf x},\tau)\rangle={\bf 0}$ and $\langle Q_i({\bf
x},\tau)Q_j({\bf x}',\tau')\rangle=\delta_{ij}\delta({\bf x}-{\bf
x}')\delta(\tau-\tau')$. We see that the only control parameters are the
dimensionless temperature $\eta=1/\theta$  and the number of particles $N$. In
the thermodynamic limit $N\rightarrow +\infty$ with fixed $\theta$, the noise
term disappears and we recover the deterministic Smoluchowski-Poisson system.

\section{The dimensionless stochastic equations of motion}
\label{sec_dimm}

If we introduce the dimensionless variables $X$, $\eta=1/\theta$, $\tau$ and $q$
through the relations
\begin{eqnarray}
 {x}=R{X},\quad \eta=\beta GMmR=\frac{1}{\theta},\nonumber\\
  t=\frac{\xi R}{GM}\tau,\quad {R}=\sqrt{\frac{GM}{\xi R}}{q},
\end{eqnarray}
we find that Eqs. (\ref{efn1}) and (\ref{efn2}) may be rewritten as
\begin{eqnarray}
\frac{d X_i}{d\tau}=\frac{1}{N}\sum_j {\rm sgn}
(X_j-X_i)+\frac{1}{2}(2X_i-1)+\sqrt{{2\theta}}q_i(\tau),\nonumber\\
\end{eqnarray}
where $q_i(\tau)$ is a Gaussian white noise 
satisfying $\langle q_i(\tau)\rangle={0}$ and $\langle
q_i(\tau)q_j(\tau')\rangle=\delta_{ij}\delta(\tau-\tau')$.

\section{The gravitational force corresponding to the modified Newtonian model in $d=1$}
\label{sec_gm}

In $d=1$, the  modified Poisson equation (\ref{efd2}) takes the form
\begin{eqnarray}
\label{gm0}
\frac{\partial^2\Phi}{\partial x^2}=2G (\rho-\overline{\rho}),
\end{eqnarray}
where $\overline{\rho}=M/(2R)$. Integrating this equation from $0$ to $x$ and using the boundary condition $\Phi'(0)=0$, we get
\begin{eqnarray}
\label{gm1}
\frac{\partial\Phi}{\partial x}=2G\int_0^x (\rho(x')-\overline{\rho})\, dx'.
\end{eqnarray}
Similarly, integrating the modified Poisson equation (\ref{gm0}) from $x$ to $R$ and using the boundary condition $\Phi'(R)=0$, we get
\begin{eqnarray}
\label{gm2}
-\frac{\partial\Phi}{\partial x}=2G\int_x^R (\rho(x')-\overline{\rho})\, dx'.
\end{eqnarray}
Subtracting  these two expressions, we obtain
\begin{eqnarray}
\label{gm3}
\frac{\partial\Phi}{\partial x}=G\int_0^x (\rho(x')-\overline{\rho})\, dx'-G\int_x^R (\rho(x')-\overline{\rho})\, dx'.\quad
\end{eqnarray}
This equation can be rewritten as
\begin{eqnarray}
\label{gm4}
\frac{\partial\Phi}{\partial x}=G\int_0^R (\rho(x')-\overline{\rho})\, {\rm sgn} (x-x')\, dx',
\end{eqnarray}
where ${\rm sgn} (x)=1$ if $x>0$ and  ${\rm sgn} (x)=-1$ if $x<0$. Therefore, the gravitational field $F=-\Phi'$ at $x$ is
\begin{eqnarray}
\label{gm5}
F(x)=G\int_0^R (\rho(x')-\overline{\rho})\, {\rm sgn} (x'-x)\, dx'.
\end{eqnarray}
It can be rewritten as
\begin{eqnarray}
\label{gm6}
F(x)=G\int_0^R \rho(x')\, {\rm sgn} (x'-x)\, dx'+G\overline{\rho}(2x-R).
\end{eqnarray}
Finally, using $\rho(x)=\sum_j m\delta(x-x_j)$, the force by unit of mass experienced by the $i$-th particle due to the interaction with the other particles and with the background density $\overline{\rho}$ is
\begin{eqnarray}
\label{gm7}
F(x_i)=Gm\sum_{x_j\in S_+} {\rm sgn} (x_j-x_i)+G\overline{\rho}(2x_i-R),
\end{eqnarray}
where $S_+$ denotes the interval $[0,R]$. We note that the first term is equal
to the mass situated on the right of the $i$-th particle ($x_j>x_i$)  minus the
mass situated on its left ($x_j<x_i$). This is a striking property of the
gravitational force in one dimension. On the other hand, the background density
$\overline{\rho}$ creates a force directed towards the wall $x=R$ when $x_i>R/2$
and towards the center of 
the domain $x=0$ when $x_i<R/2$.\footnote{This observation may
help interpreting the results of \cite{cd}. The modified Boltzmann-Poisson
equation (\ref{p5}) with the boundary conditions $\Phi'(0)=0$ and $\Phi'(R)=0$
admits an infinite
number of solutions, presenting $n$ clusters (oscillations). However, the
solutions with $n\ge 2$ are unstable. Only the solution $n=1$,
corresponding to the density profiles shown in Fig. \ref{profilesstochastique},
is stable. In that case, the particles are concentrated at $x=0$ or $x=R$. The
solution in which the particles are concentrated at $x=R/2$, corresponding to
$n=2$, is unstable. Physically, this is due to the effect of the background
density $\overline{\rho}$ in the modified Poisson equation (\ref{gm0}) that
``pushes'' the
particles either towards $x=0$ or $x=R$ according to the expression of the force
(\ref{gm7}).}

\section{Derivation of the Kramers formula from the instanton theory}
\label{sec_instanton}

In this Appendix, we calculate the escape rate $\Gamma$ of a system of Brownian
particles with long-range interactions across a barrier of free energy by using
the instanton theory. This provides a justification of the Kramers formula
(\ref{lifetime2}) giving the typical lifetime of a metastable state.

We first consider an overdamped particle moving in one dimension in a bistable
potential $V(x)$ and subject to a Gaussian white noise $R(t)$. The Langevin
equation is $\dot x=-V'(x)/\xi m+\sqrt{{2k_B T}/{\xi
m}}R(t)$.
When $T=0$ (no noise), the evolution is deterministic and the particle relaxes
to
one of the  minima of the potential since $dV/dt=-(1/\xi m)V'(x)^2\le 0$.
When $T>0$, the particle switches back and forth between the two
minima (attractors). For $T\rightarrow 0$, the transition between the two
metastable states is a rare event.
One important problem is to determine the rate $\Gamma$ for the particle,
initially located in a metastable state, to cross the potential barrier and
reach the other  metastable state. The most probable path for the stochastic
process $x(t)$ between $(x_1,t_1)$ and $(x_2,t_2)$  was first determined by
Onsager and Machlup \cite{om}. The
probability of the path $x(t)$ is $P[x(t)]\propto e^{-S[x]/k_B T}$ where
$S[x]=(\xi m/4)\int dt\, (\dot x+V'(x)/\xi m)^2$ is the Onsager-Machlup
functional. The probability to pass from  $(x_1,t_1)$ to $(x_2,t_2)$ is
$P[x_2,t_2|x_1,t_1]=\int {\cal D}x e^{-S[x]/k_B T}$. 
The functional $S[x]$ may be called an action by analogy with the path-integrals
formulation of quantum mechanics (the temperature $T$ plays the role of the
Planck constant $\hbar$ in quantum mechanics) \cite{feynman}. The most probable
path $x_c(t)$ connecting two states is called an ``instanton'' \cite{instanton}.
It is obtained by minimizing the Onsager-Machlup functional.  In the weak noise
limit $T\rightarrow 0$, the transition probability from one state to the other
is dominated by the most probable path: $P[x_2,t_2|x_1,t_1]\propto
e^{-S[x_c]/k_B T}$. Therefore, the action of the most probable path between two
metastable states determines the escape rate $\Gamma\sim {\rm
exp}(-S[x_c]/k_B T)$ of the particle over the potential barrier. One finds
$\dot x_c=+ V'(x_c)/\xi m$ for the uphill path and $\dot
x_c=- V'(x_c)/\xi m$ for the downhill path so that
$S[x_c]=\Delta V$
where $\Delta V=V_{max}-V_{meta}$ is the barrier of potential energy. This
yields the Arrhenius
law $\Gamma\sim {\rm exp}(-\Delta V/k_B T)$ stating that the transition rate
is inversely proportional to the exponential of the potential barrier. A general
path-integrals formalism determining the escape rate of a particle in the weak
noise limit
has
been developed by Bray {\it et al.} \cite{bray}. Their theory accounts for
white noises for which $S[x_c]= \Delta V$ and for 
exponentially correlated noises for which $S[x_c]\neq \Delta V$.  The
instanton
theory has been formalized by Freidlin and Wetzel \cite{fw} in relation to the
theory of large deviations \cite{touchette}. It has been applied (and
extended) to various
systems such as scalar fields described by the Ginzburg-Landau equation
\cite{bray,e}, interacting magnetic moments \cite{berkov}, nucleation
\cite{lutsko}, and two-dimensional fluid flows \cite{blz}. We apply it here to
the case of Brownian particles with long-range interactions described by the
stochastic Smoluchowski equation (\ref{sto5}).

In order to apply the formalism of instanton theory in a simple setting, it is
convenient to consider spherically symmetric distributions.  If we ignore the
noise in a first step, the mean field Smoluchowski equation (\ref{h13}) can be
written in terms of the integrated density $M(r,t)=\int_0^r \rho(r',t)S_d
{r'}^{d-1}\, dr'$ as
\begin{eqnarray}
\label{inst1}
\xi\frac{\partial M}{\partial t}=\frac{k_B T}{m}\left
(\frac{\partial^2M}{\partial r^2}-\frac{d-1}{r}\frac{\partial M}{\partial
r}\right )+\frac{\partial M}{\partial r}\frac{\partial\Phi}{\partial r}.
\end{eqnarray}
For self-gravitating systems, using the Gauss theorem $\partial\Phi/\partial
r=GM(r,t)/r^{d-1}$ in Eq. (\ref{inst1}), we see that the Smoluchowski-Poisson
system is equivalent to a single partial differential equation for $M(r,t)$
\cite{sc}. The free energy (\ref{h4}) can be written as a functional of $M(r,t)$
of the form
\begin{eqnarray}
\label{inst2}
F[M]=\frac{1}{2}\int \frac{\partial M}{\partial r}\Phi\, dr\nonumber\\
+\frac{k_B T}{m}\int \frac{\partial M}{\partial r}\ln\left (\frac{1}{NmS_d
r^{d-1}}\frac{\partial M}{\partial r}\right )\, dr\nonumber\\
-\frac{dN}{2}k_B T\ln\left (\frac{2\pi k_B T}{m}\right ).
\end{eqnarray}
Since
\begin{eqnarray}
\label{inst3}
\frac{\delta F}{\delta M}=-\frac{\partial\Phi}{\partial r}
-\frac{k_B T}{m}\frac{1}{\frac{\partial M}{\partial r}}\left
(\frac{\partial^2M}{\partial r^2}-\frac{d-1}{r}\frac{\partial M}{\partial
r}\right )
\end{eqnarray}
we can rewrite Eq. (\ref{inst1}) as
\begin{eqnarray}
\label{inst4}
\xi\frac{\partial M}{\partial t}=-\frac{\partial M}{\partial r}\frac{\delta
F}{\delta M}.
\end{eqnarray}
The $H$-theorem writes
\begin{eqnarray}
\label{inst4b}
\dot F=-\frac{1}{\xi}\int_0^{+\infty}\frac{\partial M}{\partial r}\left
(\frac{\delta F}{\delta M}\right )^2\, dr\le 0.
\end{eqnarray}

We can now introduce the noise in order to recover the canonical Boltzmann
distribution at equilibrium (see the remark following Eq. (\ref{sto7})). This
leads to the stochastic partial differential  equation
\begin{eqnarray}
\label{inst5}
\xi\frac{\partial M}{\partial t}=-\frac{\partial M}{\partial r}\frac{\delta
F}{\delta M}+\sqrt{2\xi k_B T\frac{\partial M}{\partial r}}R(r,t),
\end{eqnarray}
where $R(r,t)$ is a Gaussian white noise satisfying $\langle R(r,t)\rangle=0$
and $\langle R(r,t)R(r',t')\rangle=\delta(r-r')\delta(t-t')$. Since the noise
breaks the spherical symmetry in the stochastic Smoluchowski equation
(\ref{sto5}), this result is valid only in an average sense (it is, however,
exact
for one dimensional systems). A direct derivation of Eq. (\ref{inst5}) starting
from the stochastic Smoluchowski equation (\ref{sto5}) is given in
\cite{lutsko}. We note that Eq. (\ref{inst5}) looks similar to the stochastic
Ginzburg-Landau equation (\ref{gl}) with a mobility $\Gamma$ proportional to
$\partial M/\partial r(r,t)$. The corresponding Fokker-Planck equation
for the probability density $P[M,t]$ of the profile $M(r,t)$ at time $t$ is 
\begin{eqnarray}
\label{inst5b}
&&\xi\frac{\partial P}{\partial t}[M,t]\nonumber\\
&=&\int_0^{+\infty}
\frac{\delta}{\delta M}\left\lbrace \frac{\partial M}{\partial r}\left\lbrack
k_B T \frac{\delta}{\delta M}+\frac{\delta F}{\delta M}\right\rbrack
P[M,t]\right\rbrace\, dr.\nonumber\\
\end{eqnarray}
We can check that it relaxes towards the canonical distribution
\begin{eqnarray}
\label{inst5c}
P[M]=\frac{1}{Z(\beta)}e^{-\beta F[M]}.
\end{eqnarray}

Since the distribution of the Gaussian white noise $R(r,t)$ is
\begin{eqnarray}
\label{inst6}
P[R(r,t)]\propto e^{-\frac{1}{2}\int dt\int_0^{+\infty}R^2\, dr},
\end{eqnarray}
the probability to observe the path $M(r,t)$ between $(M_1(r),t_1)$ and
$(M_2(r),t_2)$ is given by
\begin{eqnarray}
\label{inst7}
P[M(r,t)]\propto  e^{-S[M]/k_B T}
\end{eqnarray}
with the action
\begin{eqnarray}
\label{inst8}
S[M]=\frac{1}{4\xi}\int dt\int_0^{+\infty}dr\, \frac{1}{\frac{\partial
M}{\partial r}}\left (\xi\frac{\partial M}{\partial t}+\frac{\partial
M}{\partial r}\frac{\delta F}{\delta M}\right )^2.\nonumber\\
\end{eqnarray}
This is the proper generalization of the Onsager-Machlup functional for our
problem. It can be written as $S=\int L\, dt$ where $L$ is the Lagrangian.
The probability density to find the system with the profile $M_2(r)$
at time $t_2$ given that it had the profile $M_1(r)$ at time $t_1$ is
\begin{eqnarray}
\label{inst9}
P[M_2,t_2|M_1,t_1]\propto \int {\cal D}M\,  e^{-S[M]/k_B T},
\end{eqnarray}
where the integral runs over all paths satisfying $M(r,t_1)=M_1(r)$ and
$M(r,t_2)=M_2(r)$.  For $N\rightarrow +\infty$, using the scaling of Sec.
\ref{sec_ybg}, the noise is weak so that the typical paths explored by the
system are concentrated close to the most probable path. In that case, a
steepest-descent evaluation of the path integrals is possible. We thus have to
determine the most probable path, i.e. the one that minimizes the action $S[M]$.
  The equation for the most probable path (instanton) between two metastable
states (attractors) is obtained by canceling the first order variations of the
action: $\delta S=0$. Actually, it is preferable to remark that, since the
Lagrangian does not explicitly depend on time, the Hamiltonian
\begin{eqnarray}
\label{inst9b}
H=\int_0^{+\infty} \dot M\frac{\delta L}{\delta \dot M}\, dr -L
\end{eqnarray}
is conserved (we have noted $\dot M=\partial M/\partial t$). Using Eq.
(\ref{inst8}), we get
\begin{eqnarray}
\label{inst8b}
H=\frac{1}{4\xi}\int_0^{+\infty}dr\, \frac{1}{\frac{\partial M}{\partial
r}}\left (\xi\frac{\partial M}{\partial t}+\frac{\partial M}{\partial
r}\frac{\delta F}{\delta M}\right )\nonumber\\
\times\left (\xi\frac{\partial M}{\partial t}-\frac{\partial M}{\partial
r}\frac{\delta F}{\delta M}\right ).\quad
\end{eqnarray}
Since the attractors satisfy $\partial M/\partial t=0$ and ${\delta F}/{\delta
M}=0$, the constant $H$ is equal to zero. Then, we find that the instanton
satisfies
\begin{eqnarray}
\label{inst10}
\xi\frac{\partial M}{\partial t}=\mp\frac{\partial M}{\partial r}\frac{\delta
F}{\delta M}
\end{eqnarray}
with the boundary conditions $M(r,t_1)=M_1(r)$ and $M(r,t_2)=M_2(r)$. Coming
back to the original model written in terms of the density, the instanton
equation is
\begin{eqnarray}
{\partial\rho\over\partial t}=\pm\nabla\cdot \left\lbrack\frac{\rho}{\xi}\nabla
\left (\frac{\delta F}{\delta\rho}\right )\right\rbrack.
\label{inst11}
\end{eqnarray}
We note that the most probable path corresponds to the gradient driven dynamics
(\ref{h16}) with a sign $\pm$, similarly to the one dimensional problem
recalled above \cite{om,bray}. The physical interpretation of this result is
given below.
For
$N\rightarrow +\infty$ (weak noise), the main contribution to the path integral
in Eq.
(\ref{inst9}) corresponds to the path that minimizes the action. This leads to
the large deviation result
\begin{eqnarray}
\label{inst12}
P[M_2,t_2|M_1,t_1]\propto  e^{-N s[M_c]/k_B T},
\end{eqnarray}
where we have written $S=Ns$ with $s\sim 1$. We note the analogies between Eqs.
(\ref{inst7}), (\ref{inst9}), (\ref{inst12}) and Eqs. (\ref{h5}),
(\ref{h6}), (\ref{h7}).

In the limit of weak noise, and for a stochastic process that obeys a
fluctuation-dissipation relation, the most probable path between two metastable
states must necessarily pass through the saddle point \cite{bray,vanden,lutsko}.
Once
the system reaches the saddle point it may either return to the initial
metastable state or reach the other metastable state. In the latter case, it has
crossed the barrier of free energy. The physical interpretation of Eq.
(\ref{inst11}) is the following. Starting from a metastable state, the most
probable path follows the time-reversed dynamics against the free energy
gradient up to the saddle point; beyond the saddle point, it follows the
forward-time dynamics down to the metastable state. According to Eqs.
(\ref{inst8}) and (\ref{inst10}), the action of the most
probable path corresponding to the transition from the saddle point to a
metastable state (downhill solution corresponding to Eq. (\ref{inst10}) with the
sign $-$) is zero  while the action of the most probable path  corresponding to
the transition from a metastable state to the saddle point (uphill solution
corresponding to Eq. (\ref{inst10}) with the sign $+$) is non zero. This is
expected since the descent from the saddle point to a metastable state is a
``free'' descent that does not require external noise; it thus gives the
smallest possible value of zero of the action. By contrast, the rise from a
metastable state to the saddle point requires external noise. The action for the
uphill solution is
\begin{eqnarray}
\label{inst13}
S[M_c^+]=\int dt\int_0^{+\infty}dr\, \frac{\partial M}{\partial t}\frac{\delta
F}{\delta M}\nonumber\\
=\int_0^{+\infty}dr\int_{M_{meta}}^{M_{saddle}}\, \frac{\delta F}{\delta M}\,
dM=\Delta F
\end{eqnarray}
or, equivalently,
\begin{eqnarray}
\label{inst13b}
S[M_c^+]=\frac{1}{\xi}\int dt\int_0^{+\infty}dr\, \frac{\partial M}{\partial
r}\left (\frac{\delta
F}{\delta M}\right )^2\nonumber\\
=\int dt \int_0^{+\infty}\, dr \frac{\partial M}{\partial
t}\frac{\delta F}{\delta
M}=\int dt\, \dot F=\Delta F,
\end{eqnarray}
where $\Delta F=F(M_{saddle})-F(M_{meta})$. The total action for the most
probable path connecting the metastable states is
$S_c=S[M_c^{+}]+S[M_c^{-}]=\Delta F+0=\Delta F$. It is determined solely by the
uphill path. The instanton solution gives the dominant contribution to the
transition rate for a weak noise. Therefore, the rate for the system to pass
from one metastable state to the other (escape rate) is
\begin{eqnarray}
\label{inst14}
\Gamma\sim e^{-\Delta F/k_B T}.
\end{eqnarray}
The typical lifetime of a metastable state is $\sim\Gamma^{-1}$ returning the
Arrhenius-Kramers formula (\ref{lifetime2}) stating that the transition rate is
inversely proportional to the exponential of the barrier of free energy.

It may also be interesting to discuss the link with the principle of maximum
dissipation of free energy \cite{onsager}. We introduce the dissipation
functions
\begin{eqnarray}
\label{inst15}
E_d=\frac{1}{2}\xi\int_0^{+\infty}\frac{1}{\frac{\partial M}{\partial r}}\left
(\frac{\partial M}{\partial t}\right )^2\, dr,
\end{eqnarray}
\begin{eqnarray}
\label{inst16}
E_d^*=\frac{1}{2\xi}\int_0^{+\infty}\frac{\partial M}{\partial r}\left
(\frac{\delta F}{\delta M}\right )^2\, dr.
\end{eqnarray}
This type of functionals first appeared in the works of Lord Rayleigh
\cite{lord} and Onsager \cite{onsager}.
We also recall that
\begin{eqnarray}
\label{inst17}
\dot F=\int_0^{+\infty}\frac{\delta F}{\delta M}\frac{\partial M}{\partial t}\,
dr.
\end{eqnarray}

First, we note that the mean field Smoluchowski equation (\ref{inst4}) can be
obtained by minimizing the functional $\dot F+E_d$  with respect to the variable
$\dot M$. This is the principle of maximum dissipation (in absolute value) of
free energy \cite{onsager} (see also \cite{rs} and Sec. 2.10.3 of \cite{nfp}).
Actually, in the present context, the validity of this  ``principle'' can be
proven rigorously since the mean field Smoluchowski equation can be derived from
the $N$-body dynamics of the Brownian particles when $N\rightarrow +\infty$. The
first variations $\delta(\dot F+E_d)=0$ return Eq. (\ref{inst4}). This
corresponds to a true minimum since $\delta^2(\dot
F+E_d)=\frac{1}{2}\xi\int_0^{+\infty} (\delta\dot M)^2/(\partial M/\partial r)\,
dr\ge 0$. Then, we find that $\dot F=-2E_d=-2E_d^*$.

Using Eqs. (\ref{inst15})-(\ref{inst17}), the action
(\ref{inst8}) can be expanded as
\begin{eqnarray}
\label{inst18}
S[M]=\frac{1}{2}\int (E_d+E_d^*+\dot F)\, dt.
\end{eqnarray}
This is the counterpart of Eq. (4-18) of Onsager and Machlup \cite{om}. The
minimization of $S$ for variations with respect to $\dot M$ is equivalent to
the minimization of $\dot F+E_d$ which returns the principle of maximum
dissipation of free energy and the mean field  Smoluchowski equation
(\ref{inst4}). This corresponds to the downhill instanton solution for which we
have $\dot F=-2E_d=-2E_d^*$ and $S=0$. On the other hand, for the uphill
instanton solution corresponding to the mean field Smoluchowski equation
(\ref{inst4}) with the opposite sign, we have $\dot F=2E_d=2E_d^*$ and
$S=\int \dot F\, dt=\Delta F$. Therefore, the probability for the system to
pass from a metastable state to a state $M(r)$ is $P[M]\propto e^{-\beta
S}\propto e^{-\beta (F[M]-F_{meta})}$. This is in agreement with the heuristic
arguments of Sec. \ref{sec_lifetime} that are now justified from the instanton
theory.

\end{document}